UNIVERSITY OF MIAMI

REGULATION OF MOUSE LEARNING AND MOOD BY THE ANTI-INFLAMMATORY CYTOKINE INTERLEUKIN-10

By

Ryan Joseph Worthen

A  DISSERTATION

Submitted to the Faculty
of the University of Miami
in partial fulfillment of the requirements for
the degree of Doctor of Philosophy

Coral Gables, Florida

December 2020

UNIVERSITY OF MIAMI

A dissertation submitted in partial fulfillment of
the requirements for the degree of
Doctor of Philosophy

REGULATION OF MOUSE LEARNING AND MOOD BY THE ANTI-
INFLAMMATORY CYTOKINE INTERLEUKIN-10

Ryan Joseph Worthen

Approved:

________________________
Eléonore Beurel, Ph.D.
Professor of
Psychiatry and Behavioral Sciences

________________________
Coleen Atkins, Ph.D.
Associate Professor of
Neurological Surgery

________________________
Kevin Park, Ph.D.
Professor of
Neurological Surgery

________________________
Gaofeng Wang, Ph.D.
Associate Professor of
Human Genetics

________________________
Anthony Filiano, Ph.D.
Assistant Professor of Neurosurgery
Duke University School of Medicine

________________________
Guillermo Prado, Ph.D.
Dean of the Graduate School

Major depressive disorder is a widespread mood disorder. One of the most debilitating symptoms patients often experience is cognitive impairment. Recent findings suggest that inflammation is associated with depression and impaired cognition. Pro-inflammatory cytokines are elevated in the blood of depressed patients and impair learning and memory processes, suggesting that an anti-inflammatory approach might be beneficial for both depression and cognition. Utilizing the learned helplessness paradigm, we first established a mouse model of depression in which learning and memory are impaired. We found that learned helplessness (LH) impaired novel object recognition (NOR) and spatial working memory. LH mice also exhibited reduced hippocampal dendritic spine density and increased microglial activation compared to non-shocked (NS) mice or mice that were subjected to the learned helpless paradigm but did not exhibit learned helplessness (non-learned helpless, or NLH). These effects were mediated by microglia, as treatment with PLX5622, which depletes microglia and macrophages, restored learning and memory and hippocampal dendritic spine density in LH mice. However, PLX5622 also impaired learning and memory and reduced hippocampal dendritic spine density in NLH mice, suggesting that microglia in NLH mice are involved in the production of molecules that promote learning and memory. We found that microglial interleukin (IL)-10 levels were reduced in LH mice and IL-10 administration was sufficient to restore NOR, spatial working memory, and hippocampal dendritic spine density in LH mice, and in NLH mice treated with PLX5622, consistent with a pro-cognitive role for IL-10. Altogether, these data demonstrate the critical role of IL-10 in promoting learning and memory after learned helplessness.

DEDICATION

To my mother, Paula Worthen, who taught me, from an early age, to deeply appreciate the happiness life brings and to understand that many fellow human beings are burdened with worries I might not ever have to directly face, instilling in me an intense empathy that would come to define who I am. The constant and unconditional love and support of my parents provided the indispensable foundation from which my burgeoning curiosity was free to roam unencumbered, driving my passions for science, the advancement of medicine, and the collective effort toward greater human flourishing, eventually leading me to explore my potential within the arena I find myself currently. Everything that I am, or aspire to be, is because of my parents' love.

## ACKNOWLEDGEMENT

I owe my mentor, Dr. Eléonore Beurel, a great deal of thanks. I have been enormously lucky to have a mentor who is attentive, pointedly constructive, accommodating, dependable, staunchly supportive, generous with her time and her efforts, and a pure pleasure to work with. In possession of seemingly inexhaustible energy, she somehow fulfills the responsibilities of directing research in the lab in addition to those required of a parent, has ensured a steady stream of funding, conducts benchwork on the side, and all the while has remained intimately involved in my development as a scientist. Challenges presented to me were balanced with encouragement, making my instruction just as much an exercise in self-discovery as it was the mastering of fundamental principles of research. My training under her guidance has been the most enriching and motivating experience of my professional career in science and for this she has my deep gratitude.

# TABLE OF CONTENTS

# LIST OF FIGURES

LIST OF TABLES

# Chapter 1 – Introduction and Background

## 1.1    Major depressive disorder

Major depressive disorder (MDD) is a wide-spread, debilitating, and potentially fatal, mood disorder with a lifetime prevalence of around 21% in the United States (Hasin et al., 2018) where the associated annual economic burden (e.g. lost productivity) is estimated to be around $200 billion (Bloom, 2011). MDD is a serious and uncontrollable alteration in experienced emotion and perception, clinically defined as the concurrent presentation of five or more of nine specified symptoms which are present nearly every day for a minimum of two weeks (see Table 1). Symptoms are pronounced such that clinically significant distress or impairment in social or occupational functioning is observed (DSM-5, American Psychiatric Association, 2013). Although it has been proposed that a deficiency in monoamines causes the disease, the etiology of major depression remains unknown, but is thought to involve a combination of genetic, biological, and environmental factors (Wohleb et al., 2016).

| Table 1. Major Depressive Disorder Diagnostic Criteria – DSM-5 |
| --- |
| 1. Depressed mood most of the day |
| 2. Markedly diminished interest or pleasure in all, or almost all, activities most of the day |
| 3. Significant increase or decrease in appetite, or gain or loss of weight (when not dieting) |
| 4. Insomnia or hypersomnia |
| 5. Psychomotor agitation or retardation |
| 6. Fatigue or loss of energy |
| 7. Feelings of worthlessness or excessive or inappropriate guilt (which may be delusional) |
| 8. Diminished ability to think or concentrate or indecisiveness |
| 9. Recurrent thoughts of death, suicidal ideation with or without a specific plan, or a suicide attempt |

Over the last decade, a realization has solidified within psychiatry that the historical optimism regarding the utility of serotonin modulators to treat depression may have perhaps been misplaced. Antidepressant drugs currently in clinical use, the vast majority of which act to increase synaptic concentrations of monoamines, are often inefficient as only a fraction of patients respond to any one treatment, meaning many patients undergo multiple trials with different families of antidepressants until an effective drug is found. This approach lacks both direction and precision and can understandably be frustrating for patients and clinicians alike; with treatment strategies that are still too generalized and patient response to any one treatment so unpredictable, outcomes are often left to random chance (Al-Harbi, 2012). Accordingly, because antidepressants typically take about three to four weeks to provide mood improvement, many patients stop taking their antidepressant medication before experiencing the potential benefit, which contributes to the high rate of relapse (Burcusa & Iacono, 2007).

The STAR*D trial (completed in 2006), the largest and most comprehensive clinical assessment of antidepressant efficacy to date, measured patient response to the selective serotonin reuptake inhibitor (SSRI) citalopram plus nine additional antidepressants to either add in combination or switch to as monotherapy, providing us with the best indication available of overall treatment effectiveness. Study investigators concluded (contentiously) that 28-33% of patients experienced remission of symptoms with citalopram and 67% of patients eventually experienced remission after trialing different drug combinations, leaving one-third of patients for whom no treatment could offer relief. However, this bleak picture is made all the more disheartening when considering: 1) according to study investigators, the 67% cumulative remission rate was a theoretical value calculated with the assumption that study dropouts would have had the same remission rate as those who stayed in the study; when this calculation is performed using the actual remission rates of study dropouts, the data yields an overall remission rate of 46% (Pigott, 2015) and 2) there is now convincing evidence consistent across a decade's worth of antidepressant meta-analyses that an extraordinarily large placebo effect can account for most, if not all, of the measured response to antidepressant drugs (Kirsch, 2019).

As our theoretical understanding of the neurobiological complexity of depression continues to expand, and the number of depressed patients grows, the scope of underlying mechanisms hoped to be amenable to therapeutic intervention has also grown. This shift in thinking has most recently been brought to the fore with the remarkable success of ketamine, an N-methyl-D-aspartate (NMDA) receptor antagonist that acts on the glutamatergic system, as the first FDA-approved rapid-acting antidepressant for the treatment of suicidality in severely depressed patients.

Modern brain imaging techniques, including functional magnetic resonance imaging (fMRI) and neurotransmitter receptor binding studies, are revealing that the symptoms of MDD may become manifest through a variety of underlying and interconnected disease processes, possibly involving dysfunction in dopaminergic, noradrenergic, GABAergic, and/or glutamatergic neural circuits, in addition to well-studied serotonergic pathology (Belujon et al., 2016). Thus, the rising consensus is that MDD is likely comprised of a broad range of disease subtypes, each requiring knowledge of the unique neuropathophysiology leading to the set of symptoms being presented in order to develop more efficacious treatment strategies tailored to the patient.

### 1.2    Cognitive deficits associated with depression

A symptom that is both commonly experienced by depressed patients and measured to be particularly debilitating is impairment of cognition. Depressed patients often experience attention, concentration, perception, executive function, and processing speed deficits, in addition to depressed mood, that impede everyday functions (Salehinejad et al., 2017; Zuckerman et al., 2018). Evidence suggests that these cognitive deficits can persist beyond the acute depressive episode, and are often not ameliorated by antidepressant treatment (Bora et al., 2013; Gonda et al., 2015). It is possible that some of these cognitive impairments might be the result of anatomical changes that occur in the brains of depressed patients, as was shown in seminal studies using MRI (Sheline et al., 2003). Neuroimaging studies have consistently demonstrated that MDD is associated with structural and functional abnormalities in the prefrontal cortex (PFC) (Salehinejad et al., 2017), a brain region that is heavily involved in cognitive control, working memory, and attention control governing goal-directed behavior (Gazzaniga, 2014; Miller & Cohen, 2001). It has also been well-established that depressed patients display significant reductions in the size of the hippocampus (Bremner et al., 2000). Owing to the critical importance of the hippocampus for proper functioning of learning and memory processes (Sapolsky, 2001), these anatomical changes have been the presumed proximal cause of cognitive impairments measured in depressed patients. Because hippocampal volume is negatively correlated with the number of days of untreated depression (Sheline et al., 2003), cognitive impairments observed within the first episode of the disease still require explanation. However, as will be covered in detail in subsequent sections, evidence of inflammatory processes in some depressed patients suggests a plausible explanation for the link between MDD and cognitive deficits, theoretical consequences of which include suppression of neurite outgrowth, inhibition of neuron repair, and synaptic weakening, and further suggests that disruptions in emotion regulation and impairment of learning and memory share common pathophysiological origins.

Beck's cognitive model of depression developed over 50 years ago, postulated that the development and maintenance of depression relies on biased acquisition and processing of information (Beck, 1967). According to this model, cognitive impairment associated with depression has the potential to synergistically worsen the disease through the establishment and reinforcement of maladaptive thought patterns, or schemas, leading to negatively biased attention, information processing, memory, and rumination, suggesting that negative cognitive bias plays a critical and insidious role in depressive pathology. The functional and neurobiological architecture of Beck's model has been identified, involving in part the limbic system (Disner et al., 2011). Furthermore, the effectiveness of cognitive behavioral therapy (CBT) in the treatment of depression, whereby the patient is guided by the therapist to strategically uncover and target these maladaptive schemas, lends support to the theory's validity, at least in the subset of patients for whom CBT is beneficial. However, meaningful characterization of the neurobiological pathogenesis of cognitive impairment in MDD, as well as explanations for its persistence in treatment-resistant cases, remains to be realized (Gałecki & Talarowska, 2017).

### 1.3    Learned helplessness and animal models of depression

Despite the inherent challenge of replicating the largely subjective assortment of symptoms associated with MDD in non-human animals, several models of depression have been developed to study disease mechanisms and investigate novel therapeutic strategies.

| Table 2. Preclinical approaches to modelling aspects of depressive symptomatology in rodents | |
| --- | --- |
| **Behavioral model** | **Depressive-like symptom(s) with theoretical MDD counterpart(s)** |
| Chronic mild stress (CMS)<br><br>and<br><br>Chronic unpredictable stress (CUS) | • anhedonia-like symptoms:<br>  □ reduced preference for sucrose<br>• hippocampal volume reduction<br>• HPA-axis activation<br>• immune activation/elevated plasma cytokines<br>• self-grooming behaviors reduced<br>• sensitivity to antidepressant treatments |
| Corticosterone manipulation | • HPA-axis activation |
| Drug withdrawal | • anhedonia-like symptoms:<br>  □ reduced preference for sucrose<br>• drug-seeking behavior |
| Early life stressors<br>(ex. maternal separation) | • HPA-axis activation<br>• hippocampal volume reduction<br>• immune activation/elevated plasma cytokines |
| Forced swim test (FST) | • motor retardation (diminished pro-active, self-protective behavior)<br>• sensitivity to acute and chronic antidepressant treatments |
| Genetic modifications | • increased sensitivity to stress<br>• learning and memory impairments<br>• metabolic alterations<br>• wide variety of potential applications |
| Immunogenic injection (ex. LPS) | • endocrine disruptions<br>• immune activation/elevated plasma cytokines<br>• sickness behavior |
| Learned helplessness paradigm (LH)<br>(subjection to unpredictable and uncontrollable aversive stimulus, e.g. foot-shock) | • anhedonia-like symptoms:<br>  □ diminished pursuit of sex<br>  □ reduced preference for sucrose<br>• circadian rhythm disturbance<br>• HPA-axis activation<br>• immune activation/elevated plasma cytokines<br>• learned helplessness (phenomenological syndrome; passivity in face of aversive stimulus)<br>• learning and memory impairments<br>• motor retardation (diminished pro-active, self-protective behavior)<br>• sensitivity to short-term antidepressant treatments<br>• weight loss |
| Olfactory bulbectomy | • hippocampal volume reduction<br>• hyperactivity in novel environment, passive avoidance deficits |
| Social defeat stress | • metabolic syndrome (including insulin resistance)<br>• sensitivity to antidepressant treatments<br>• social withdrawal<br>• weight gain |
| Tail suspension test (TST) | • motor retardation (diminished pro-active, self-protective behavior)<br>• sensitivity to acute and chronic antidepressant treatments |

[Table 2 outlines the widely used preclinical behavioral paradigms in rodents and the corresponding depressive symptoms that are modeled (Abelaira et al., 2013; Cryan & Mombereau, 2004; Gururajan et al., 2019; Nestler & Hyman, 2010)] Much of the progress in animal modeling has centered on the observation that stress as well as psychosocial losses are potent precipitators of both behavioral and physiological changes that mimic those observed in depressed humans. For example, chronic social defeat stress, which involves subjecting rodents to repeated encounters with dominant conspecific rivals, results in subordinated animals displaying a number of depressive-like symptoms including social withdrawal as well as metabolic syndrome characterized by insulin resistance and weight gain, both of which can be effectively treated with antidepressants (Krishnan et al., 2007). Additionally, purely physical stressors (i.e. without a social component) have also been successfully employed to precipitate depressive-like states in rodents. For instance, chronic mild and chronic unpredictable stress both involve subjecting rodents to repeated stressors such as restraint, electric foot- or tail-shock, or sleep disruption over a period of weeks and have been shown to elicit anhedonia-like behavior including reduced preference for sucrose (Nestler & Hyman, 2010). Symptoms induced by the chronic stress paradigms can also be reversed following treatment with antidepressant medications (Willner, 2005).

Supported by a theoretical framework derived from Beck's cognitive model of depression (Vollmayr & Gass, 2013), one of the most well-validated models of depression is the learned helplessness paradigm, which subjects animals to a series of uncontrollable, unpredictable, and inescapable foot-shocks over a period of hours resulting in a subset of susceptible animals becoming learned helpless, such that when re-exposed to foot-shocks that are escapable, the animals display diminished to absent motivation to avoid the shocks and fail to escape (Seligman, 1975). Rather than being a learned behavior per se, the passivity of learned helplessness is understood to be the default, unlearned response to prolonged aversive events, mediated by reactive changes in neural circuitry (Maier & Seligman, 2016) and has been demonstrated to exist in humans as well as many other mammals (Enkel et al., 2010; Everaert et al., 2012; Richter et al., 2012). Considered a hallmark of depressive-like behavior in rodents, learned helplessness has been demonstrated to include symptoms such as weight loss, altered sleep patterns, reduced sexual activity, hypothalamic-pituitary-adrenal (HPA) axis activation, as well as learning and memory impairments, representing instantiations of a large number of symptoms experienced by depressed patients (Nestler & Hyman, 2010; Takamori et al., 2001). Additionally, learned helpless symptomatology responds to antidepressant treatments, affording the learned helplessness paradigm excellent face and predictive validity as a depression model, and making it an important tool in the investigation into depressive pathophysiology and novel treatment approaches (Abelaira et al., 2013)

**Chapter 2 – Inflammation: Target for Therapy**

**2.1 The inflammatory theory of depression and the effect of neuroinflammation on brain function**

One of the new avenues being explored in depression research focuses on the role of inflammation as a potential pathogenic element. Similar to the immune response to traumatic injury and other physical stress to the body, psychosocial stress induces an inflammatory response within the central nervous system (CNS), termed neuroinflammation. Mediated by the production of pro-inflammatory cytokines and other secondary messengers, neuroinflammation activates responsive CNS and peripheral cells, which adopt modes of cellular operation that are meant to be protective. This physiological stress response may have been selected evolutionarily to confer a protective advantage to the body at a time when physical injury was more tightly linked to psychological stress (Benarroch, 2013; Le Thuc et al., 2015; Miller & Raison, 2016; Ransohoff et al., 2015; Rivest, 2009; Shatz, 2009). If properly regulated, this immune activation is expected to be transient, facilitating resolution to the putative insult and serving to protect the brain (Bitzer-Quintero & González-Burgos, 2012; Le Thuc et al., 2015). When/if the CNS suffers trauma, circulating immune cells, such as monocytes and lymphocytes (Stoll & Jander, 2005; Whitney et al., 2009), are recruited from the periphery to join CNS-resident microglia (Bitzer-Quintero & González-Burgos, 2012), and are guided to the site of CNS injury via chemokines, molecular homing signals (Ransohoff & Engelhardt, 2012). Upon encountering damaged tissue, recruited cells act to repair the insult by phagocytosing apoptotic cellular debris and steadily increasing the release of anti-inflammatory cytokines, which dampen inflammation, as well as neurotrophins, growth factors that promote neurogenesis and synaptogenesis, driving recovery after neuronal damage and culminating in resolution of neuroinflammation (Bitzer-Quintero & González-Burgos, 2012; Le Thuc et al., 2015; Mathieu et al., 2010). Crucially, to promote health, immune cell activity must be tightly constrained both spatially and temporally, as neurons, whose proper function is important for an organism's well-being, are highly sensitive to their physiological environment.

The inflammatory theory of depression proposes that an inappropriately elevated or extended immune response to stress has the potential to initiate a cascading activation of immune-responsive cells, a process that results in perpetuation and amplification of inflammatory signaling and disruption of neuronal communication in circuits relevant to emotional state such that mood becomes dysregulated. Stress is thought to release damage-associated molecular patterns (DAMPs) such as ATP, heat shock proteins, high mobility group box 1 (HMGB1), glucose, and uric acid as well as pathogen-associated molecular patterns (PAMPs) such as flagellin and the bacterial endotoxin lipopolysaccharide (LPS) associated with microbes escaping an increasingly permeable gut (Felger & Lotrich, 2013). These "danger" molecular signals are recognized by pattern recognition receptors, expressed by circulating peripherally-derived immune cells, which activate the "nuclear factor kappa-light-chain-enhancer of activated B cells" (NF-κB) pathway and "NLR (nucleotide-binding oligomerization domain (NOD)-like receptor) family pyrin domain containing 3" (NLRP3) inflammasome, leading to increased production of pro-inflammatory cytokines, in turn promoting activation of both astrocytes and microglia, whose own pro-inflammatory cytokine production feeds back, amplifies, and prolongs neuroinflammation (Cox et al., 2014; Johnson et al., 2019). Chronic stress and concurrent neuroendocrine activation via the HPA axis results in prolonged exposure, and compensatory resistance, to glucocorticoids, effectively reducing immune cell sensitivity to regulatory anti-inflammatory feedback (Wohleb et al., 2016).

At a molecular level, excessively high inflammation is thought to lead to decreased neurotransmitter metabolism, increased production of neurotoxic reactive oxygen species (Whitney et al., 2009), increased glutamate excitotoxicity (Raison & Miller, 2013), inhibition of long-term potentiation (LTP), changes in dendritic spine density (Bilbo & Schwarz, 2009), and decreased production and release of brain-derived neurotrophic factor (BDNF) by neurons and astrocytes, all of which impedes synaptic plasticity and synaptic scaling (Chugh et al., 2013; Green & Nolan, 2014; Jakubs et al., 2008; Kohman & Rhodes, 2013; McAfoose & Baune, 2009; Vezzani & Viviani, 2015; Wood et al., 2011), hinders neurogenesis in the dentate gyrus of the hippocampus (Jo et al., 2015; Sapolsky, 2001), and disrupts neuronal functional integration (Jakubs et al., 2008; Wood et al., 2011) (Fig. 1). Additionally, neuroinflammation increases blood-brain barrier (BBB) permeability (Pan et al., 2011) allowing inappropriate peripheral immune cell infiltration into the CNS (Erickson & Banks, 2018), and perpetuates further production of pro-inflammatory cytokines (Le Thuc et al., 2015). Altogether, these biomolecular changes cause damage to neurons, impede healthy synaptic connectivity, and prevent repair to neural networks in both the developing (Harry & Kraft, 2012) and adult brain (Das & Basu, 2008). Depending on the specific neural circuits affected and degree of disruption, plausible consequences are believed to include depressed mood (Borsini et al., 2015) as well as impairments to learning and memory (Raison et al., 2006) (Fig. 1).

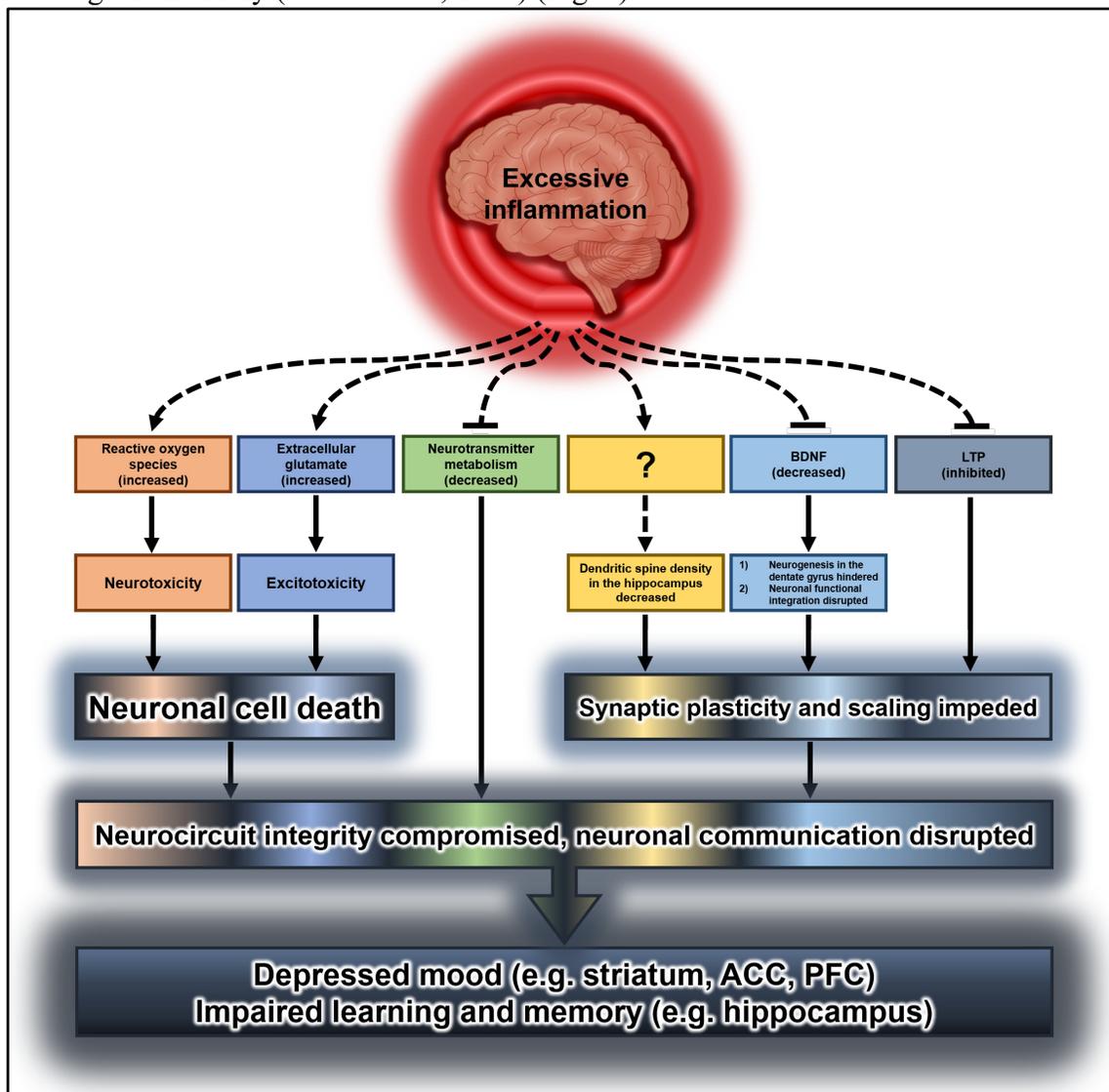

**Figure 1. Schematic summary of proposed pathophysiological consequences of dysregulated neuroinflammation.**

In support of this theoretical framing, it is well-established that the immune systems of some depressed patients display signs of heightened activity, which has been associated with symptom severity and type (Beurel et al., 2020). Many depressed patients exhibit elevated blood levels of pro-inflammatory cytokines such as IL-6 and tumor necrosis factor α (TNFα), as well as inflammatory markers like C-reactive protein (Dantzer, 2012; Raison et al., 2006). Furthermore, there is evidence that pro-inflammatory cytokines have the potential to induce depressive states and exacerbate depression severity. For instance, 45% of patients receiving interferon-alpha, an inflammatory cytokine used to treat hepatitis C, develop depressive symptoms *de novo* (Zdilar et al., 2000). Additionally, healthy volunteers administered with LPS, which triggers the production of pro-inflammatory cytokines, experience a sickness behavior characterized by depressive symptoms, and there is a direct correlation between depressed mood symptoms and measured plasma levels of TNFα (Reichenberg et al., 2001), suggesting that an increase in levels of pro-inflammatory cytokines is sufficient to modulate mood.

## 2.2 Relationship between the immune system and cognition

Neurotoxicity and disruptions in neural connectivity present obvious threats to proper cognitive function. However, current theory posits that neural activity, especially that which underpins learning and memory, is sensitive even to non-pathological changes in levels of pro- and anti-inflammatory cytokines, which mediate cell-to-cell communication, in the brain (Shatz, 2009). These immune signal mediators play key roles in the maintenance of a homeostatic balance important for the regulation of synaptic architecture (Benarroch, 2013). A testament to the delicate nature of this equilibrium, there is evidence that the pro-inflammatory cytokines IL-1β, IL-6, and TNF are involved in the preservation of a balanced neuroprotective environment in the developing brain. Similarly, mice with suppressed or absent adaptive immunity have been shown to exhibit impairments in cognition (Brynskikh et al., 2008) (see Fig. 2).

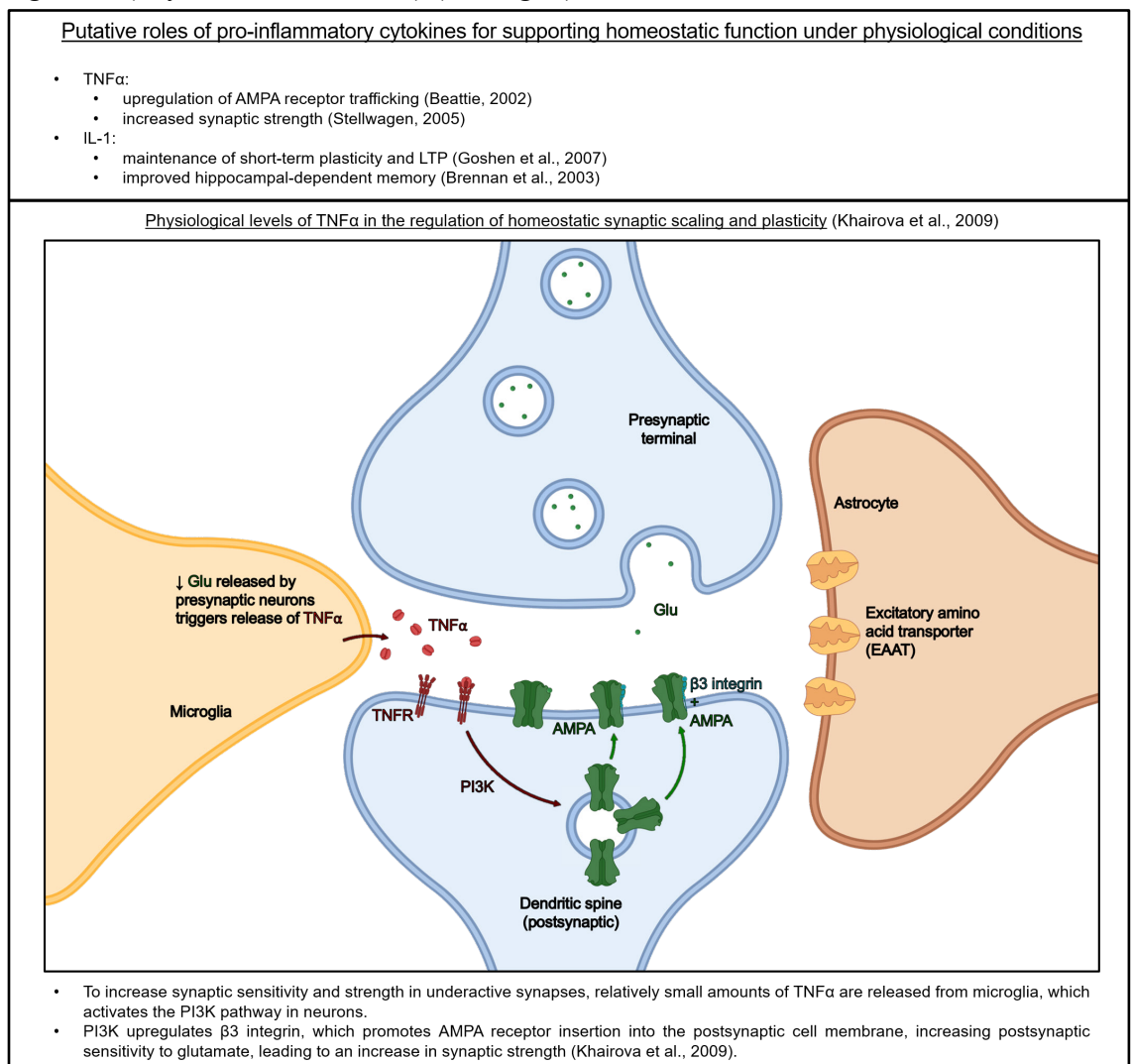

Figure 2. The putative role of selected pro-inflammatory cytokines in maintaining homeostasis under healthy conditions.

However, as mentioned previously, high levels of inflammation are severely detrimental, impairing neurogenesis, synaptic plasticity, and synaptic scaling (McAfoose & Baune, 2009). In the adult brain, pro-inflammatory cytokines exert a negative effect on cognition when they become elevated, impairing hippocampal-dependent memory (Marin & Kipnis, 2013; Yirmiya & Goshen, 2011) and promoting sickness behavior, depression, and stress in humans (Dantzer et al., 2008) (see Fig. 3).

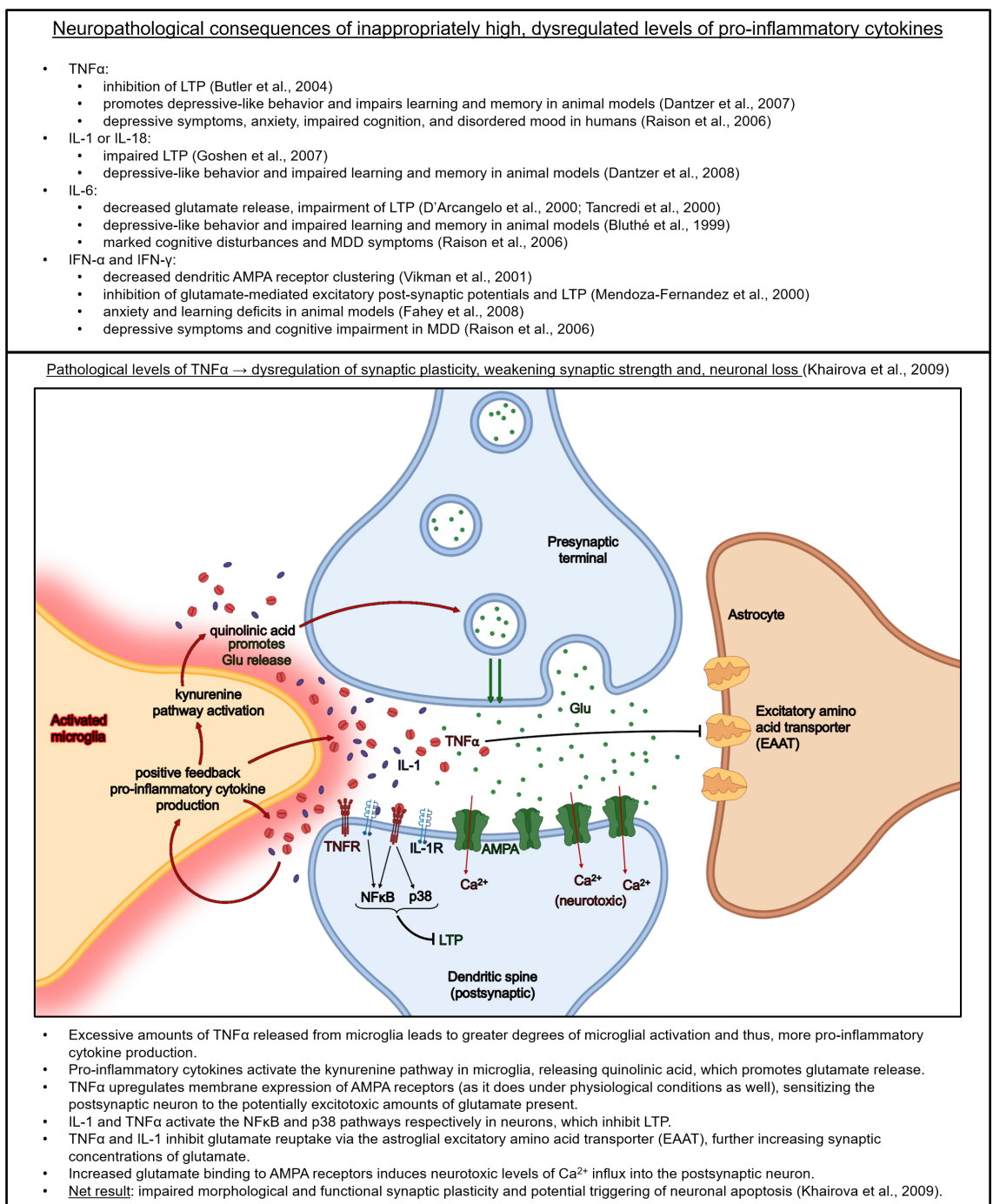

Figure 3. The pathophysiological roles of pro-inflammatory cytokines in mood disorders.

## 2.3    The anti-inflammatory cytokine interleukin-10

Conversely, anti-inflammatory cytokines, such as interleukin (IL)-10, which terminate the pro-inflammatory cascade, have been shown to be reduced in depressed patients, and to provide pro-cognitive and antidepressant actions. In the periphery, IL-10 is primarily produced by monocytes, and, to a lesser extent, by T cells and B cells (Said et al., 2010). In the CNS, IL-10 is produced by microglia and astrocytes (Park et al., 2006) upon activation of the toll-like receptor (TLR)/NF-kB pathway (Jack et al., 2005; Ledeboer et al., 2002). IL-10 is a homodimer, each of its subunits is 178 amino acids long (~21 kDa) and it acts in a paracrine or autocrine manner, as IL-10 receptors are expressed at the surface of neurons, microglia, and, to a lesser extent, astrocytes (Lobo-Silva et al., 2016). Binding of IL-10 to its heterotetrameric receptor, IL-10R, induces signaling via the Janus kinase/signal transducers and activators of transcription (JAK/STAT) signaling pathway, in particular JAK1/STAT3 (Murray, 2006), and leads to an anti-inflammatory response by promoting transcription of genes that counteract the production of pro-inflammatory cytokines (Moore et al., 2001; Murray, 2006) (Fig. 4). In neurons, IL-10 receptor signaling has been associated with: increased cell survival (Lobo-Silva et al., 2016; Zhou et al., 2009a, 2009b), producing factors that protect neurons from glutamate-induced excitotoxicity (Kwilasz et al., 2015), the regulation of adult neurogenesis (Pereira et al., 2015; Perez-Asensio et al., 2013), and the promotion of neurite outgrowth and synapse formation (Chen et al., 2016).

IL-10 has been shown to alleviate the deleterious effects of cytokines on memory and plasticity (Lynch et al., 2004; Richwine et al., 2009), blocking for example the detrimental effects of LPS on LTP (Kavanagh et al., 2004; Kelly et al., 2001; Lynch et al., 2004; Nolan et al., 2005), and rescuing learning and memory deficits in inflammation-dependent models of Alzheimer's disease (Kawahara et al., 2012; Kiyota et al., 2012; Kiyota et al., 2010). Furthermore, a higher circulating IL-10 level has been associated with lower levels of self-reported stress as well as higher measures of cognitive function (Jung et al., 2019). Consistent with this, IL-10 levels are reduced in MDD patients (Blume et al., 2011; Dhabhar et al., 2009) although not all (Syed et al., 2018) and IL-10 attenuates LPS-induced depressive-like behaviors in rats (Bluthé et al., 1999), whereas antidepressants increase IL-10 levels (Köhler et al., 2018; Kubera et al., 2001; Syed et al., 2018). Taken together, IL-10 may have beneficial actions both within cognition and depression.

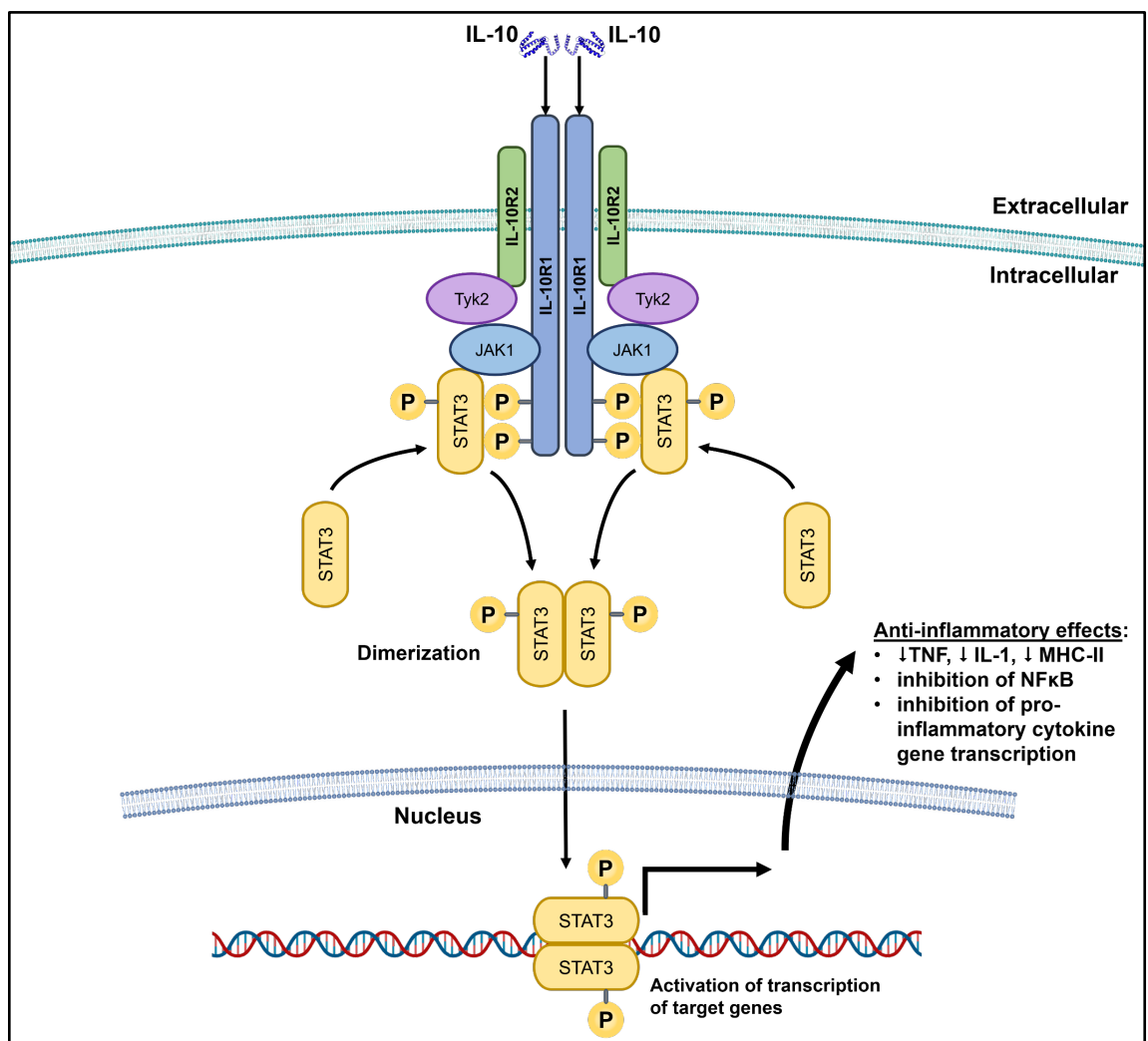

**Figure 4. Schematic overview of the IL-10R signaling cascade and important cellular effects brought about by IL-10.**

## 2.4    Role of microglia in MDD and in mouse models of depression

Microglia demonstrating markers of activation have been observed in various brain regions of MDD patients, including in the dorsolateral (dl)PFC, anterior cingulate cortex (ACC), mediodorsal thalamus, insula and hippocampus, both in post-mortem tissue and by positron emission tomography (PET) imaging (Schnieder et al., 2014; Setiawan et al., 2015; Steiner et al., 2008), and have been associated with depressive-like behaviors in rodents [review (Wohleb et al., 2015)]. Activation of microglia often is the consequence of increased inflammation in the brain (Benarroch, 2013; Dantzer et al., 2008; Delpech et al., 2015; Haroon et al., 2017; Jo et al., 2015; Ménard et al., 2016; Miller & Raison, 2016; Ransohoff et al., 2015; Rivest, 2009; Slavich & Irwin, 2014; Wachholz et al., 2016). Characterization of microglia in mouse models of depression showed that microglial cells adopt an activated phenotype expressing surface markers such as major histocompatibility complex (MHC) class II, CD86 or CD54 (Wachholz et al., 2016) and producing pro-inflammatory cytokines [e.g., IL-1β, IL-6, TNFα] (Chabry et al., 2015), toxic molecules (e.g., nitric oxide) or extracellular vesicles, such as exosomes, which might be responsible for propagating inflammatory signals throughout the brain (Frühbeis et al., 2013).

Microglial cells express several inflammatory pathways that are responsible for cytokine production, including TLR pathways. Microglial cells are a primary source of cytokines in the brain after exposure to stress (Hanisch, 2002; Kim & Joh, 2006). In response to stress, microglial cells are activated by molecules called alarmins, danger-associated molecular pattern molecules, or damage-associated molecular pattern molecules. For example, HMGB1 is up-regulated and activates TLR4 and receptor for advanced glycation end-products (RAGE) pathways resulting in increased production of proinflammatory cytokines (Franklin et al., 2014; Weber et al., 2015). Central administration of HMGB1 also induces depressive-like behaviors (Cheng et al., 2016; Franklin et al., 2014; Wu et al., 2015), altogether suggesting that up-regulation of alarmins such as HMGB1 contribute to the development of depressive-like symptoms by activating microglial cytokine production. Additionally, cytokines such as IL-6 or TNF also can disrupt BBB permeability, raising the possibility that increased BBB permeability caused by microglia-derived cytokines contributes to depression (Cheng et al., 2018; Ménard et al., 2016) by promoting the infiltration of peripheral inflammatory molecules and immune cells (Wohleb et al., 2011; Wohleb et al., 2013).

Microglial cells contribute to synaptic pruning by phagocytosis of synapses during development but also possibly during adulthood (Kettenmann et al., 2013; Miyamoto et al., 2013; Wake et al., 2013; Wohleb et al., 2018). Thus, microglia modulate neuronal circuits (Wake et al., 2013) to increase the production of neurotrophic factors in response to neuronal stimulus (Nakajima et al., 2007). Therefore, microglia activation has direct effects on neuronal function. Microglia activation impedes microglia-neuron communication via the CX3CR1-fractaline pathway after chronic stress (Corona et al., 2010; Milior et al., 2016). Furthermore, stress-induced anxiety- and depressive-like behaviors are associated with increases in neuronal colony stimulating factor 1 (CSF1) which activates microglial phagocytosis of synaptic elements in medial PFC (mPFC). This demonstrates the importance of neuron-microglia communication, as dysregulated active microglia can potentially reshape neuronal connectivity in ways that are harmful and conducive to the development of depressive-like behaviors (Wohleb et al., 2018). Moreover, microglial activation associated with stress-induced depressive-like behavior induces the activation of the kynurenic acid pathway (Jo et al., 2015), leading to the depletion of monoamines and the production of quinolinic acid (Steiner et al., 2011), and ultimately to the dysregulation of monoaminergic and glutamatergic circuits (Jo et al., 2015). Consistent with the observation that microglial cells are found in relatively high numbers in the hippocampus (Bachiller et al., 2018), a brain region relevant for stress response and neuroplasticity, the role of microglia appears to be critical for maintaining neuronal homeostasis and neurotransmission, whereas activation of microglia by stress alters this equilibrium, which can promote susceptibility to depressive-like behaviors.

The alteration of neural circuitry is thought to contribute to depressive symptoms. Because gliosis is known to be associated with alterations in neuronal connections, gliosis occurring within the hippocampus may disrupt functions for which this brain region is important, especially learning and memory. For example, neuroimaging studies definitively show that inflammation targets brain regions thought to be important in the pathophysiology of depression, the ACC and basal ganglia (Raison & Miller, 2013), both constituents of the cortico-limbic-striatal circuit, which also includes the hippocampus and amygdala. Alterations to the functioning of this circuit are likely to affect motivation as well as learning and memory (Foerde & Shohamy, 2011).

It is critical to recognize that while microglial hyperactivity resulting from dysregulation can indeed result in pathological outcomes, the protective role played by properly regulated microglia in response to immune activation is equally as consequential and is pivotal to determining the fate of overall neural function. In contrast to signs of microglial hyperactivity and neuronal damage in mPFC after a chronic stress paradigm lasting 2 weeks (Wohleb et al., 2018), exposure of mice to much longer-duration unpredictable stress over a period of 5 weeks was associated with a decline in the number of microglia in the hippocampus compared to baseline as well as depressive-like behavior, which was alleviated by microglia-activating, pro-inflammatory treatments (Kreisel et al., 2014). This suggests that significant departures in either direction of microglia from their homeostatic state (i.e. excessive activity following acute activation as well as too little activity following depletion) can be detrimental for brain function. Acute microglial depletion with the drug PLX5622 was shown to exacerbate the brain response to treatment with LPS in rodents as measured by excessive elevation in pro-inflammatory cytokines as well as motor activity consistent with sickness behavior (Vichaya et al., 2020). This suggests that microglia produce essential protective factors that are critical for healthy brain function after immunogenic insult and is a testament not only to the importance of strict regulatory constraints placed on microglial activity but also to how deeply interwoven microglia are within the complex fabric of neurobiological cellular dynamics.

During homeostasis, microglial cells play an important role in neuroplasticity, because of their crucial role in clearing apoptotic newborn neurons or dendritic spines through phagocytosis (Sierra et al., 2010; Wohleb et al., 2018), and in secreting factors that are required for the proliferation, differentiation and migration of newborn neurons (Aarum et al., 2003; Battista et al., 2006; Butovsky et al., 2006; Walton et al., 2006), promoting neurogenesis, which has been reported to be impaired in rodent models of depression (Miller & Hen, 2015). Furthermore, environmental enrichment, which confers resilience to stress, and provides antidepressant effects, induces a mild microglial activation, similarly to LPS, macrophage colony-stimulating factor (M-CSF) or granulocyte-macrophage colony-stimulating factor (GM-CSF), suggesting that the degree of microglial activation might be critical to determine the outcomes on functionality of neuroplastic circuits. Finally, minocycline administration, commonly used to block microglial activation, blocks depressive-like behavior in mice and ameliorates depressive symptoms in humans (Kreisel et al., 2014; Reis et al., 2019), and this was accompanied by a decrease in inflammatory cytokine production.

### 2.5    A theorized role for IL-10 as therapy and strategies for testing in mice

Learning is thought to involve the formation of new dendritic spines, and memory most likely involves the maturation of those spines, reinforcing newly formed synaptic connections, and integration within the broader functional network of neurons. While pro-inflammatory cytokines are known to activate glial cells leading to the removal of dendritic spines, anti-inflammatory cytokines, which attenuate gliosis, might preserve synapses, and promote the production of neuroprotective agents, such as BDNF. Little is known about the possible influence of IL-10 on dendritic spine dynamics and synaptic plasticity, processes critical for mood as well as learning and memory (Duman & Duman, 2015; Hajszan et al., 2009; Yang et al., 2015). In stress-induced depressive-like behavior, dendritic spine density is reduced, specifically in the hippocampus (Shirayama et al., 2002). Conversely, antidepressants increase dendritic spine density in rodent models of depression. Furthermore, dendritic spine architecture is regulated by gliosis. In response to stress or other insults to the brain, resident glial cells become activated, particularly microglia. It is thought that stress induces the release of DAMPs that activate the NF-κB and NLRP3 inflammasome pathways, leading to increased pro-inflammatory cytokines, which promote activation of both astrocytes and microglia. Once activated, microglia become motile and phagocytic with the demonstrated potential to engulf dendritic spines (Wohleb et al., 2018) potentially altering neuronal connectivity. Microglia are known to be capable of responding to IL-10 signaling as they express IL-10 receptors on their cell surface (Lobo-Silva et al., 2016). However, the role IL-10 plays in modulating microglial activity during the brain's stress response is not known. Neither is it known if differences in IL-10 signaling contribute to the pathogenesis of neural dysfunction, or associated cognitive impairments, in the context of depression.

Evidence for the influence of immune signaling in depression includes the observations that 1) pro-inflammatory cytokines are elevated in the blood of depressed patients and 2) the actions of antidepressants are hindered by pro-inflammatory cytokines as patients who become resistant to antidepressant therapy demonstrate higher levels of pro-inflammatory cytokines and CRP than patients who respond to antidepressants (Felger & Lotrich, 2013). This hindrance can be alleviated by anti-inflammatory drugs. A clinical trial of infliximab, an antibody targeting TNF, resulted in a modest reduction in HAM-D17 scores in MDD patients with CRP levels greater than 5 mg/L (Raison et al., 2013). Though modest, positive results like these are tantalizing - if therapies can be designed that target not just one pro-inflammatory cytokine, like TNF, but instead act to counteract and resolve inflammation generally, the benefit may be enhanced. Utilizing signaling molecules already well-known to perform this regulatory role, like anti-inflammatory cytokines, could provide beneficial outcomes in the treatment of both depressed mood and associated impairment of cognition in patients exhibiting high levels of inflammation.

The most widely studied and thoroughly characterized cytokine with anti-inflammatory roles, IL-4, has demonstrated neuroprotective and pro-cognitive roles which may portend a number of the specific functions of IL-10, if there is significant overlap in how these two cytokines are utilized biomolecularly (Marin & Kipnis, 2013). Produced by T cells, IL-4 is critical for learning and memory processes as mice without IL-4 demonstrate impaired spatial learning, which is restored upon reintroduction of IL-4 (Derecki et al., 2010). The role IL-10 plays in cognition, however, is largely unknown, although it has been suggested that IL-10 is pro-cognitive. It was shown to be critical for psychomotor coordination and hippocampal working memory during peripheral infection, as deficits in both processes were exacerbated in IL-10 knockout mice in response to LPS (Krzyszton et al., 2008; Richwine et al., 2009). Additionally, IL-10 was found to suppress activation of both microglia and astrocytes and restored neurogenesis and spatial learning of amyloid precursor protein + presenilin1 (APP+PS1) Alzheimer's mice (Kiyota et al., 2012). The well-established pro-cognitive effects of IL-4 (Gadani et al., 2012), lend support to the idea that IL-10 might be pro-cognitive.

In rodents, strategies aimed at increasing physiological levels of IL-10 have generally achieved reduction in disease symptoms and associated inflammation (Kwilasz et al., 2015). Various methodologies have been used to increase circulating IL-10 concentration, such as direct injection of the recombinant protein, viral vectors, delivery of naked plasmid, or plasmid encased in microparticles. However, there are serious drawbacks to each of these methods both in terms of their practical employment as well as potential confounding variables that are unavoidably introduced. For instance, if one is interested in measuring the anti-inflammatory properties of a given therapy, it would be ill-advised to use a delivery method that itself causes inflammation. Additionally, because our focus centered on pathological processes within the CNS, we sought a methodology that would minimize, as much as possible, influences coming from the periphery, so that a more precise picture of IL-10's role in regulating *neuro*inflammation specifically could be ascertained.

Because direct delivery via intracranial injection itself causes inflammation, we decided to test the viability of an intranasal delivery method, which had been shown previously to successfully deliver small peptides directly to the brain (Chauhan & Chauhan, 2015; Lochhead & Thorne, 2012; Mittal et al., 2014; Renner et al., 2012). Drugs administered intranasally make direct contact with the olfactory epithelium and subsequently arrive within brain parenchyma by either slow transport inside olfactory neurons extending from the olfactory bulb or via a more rapid transfer along the perineural space surrounding olfactory neurons, diffusing into the cerebrospinal fluid that surrounds the olfactory bulbs (Illum, 2004; Lochhead & Thorne, 2012; Mathison et al., 1998). The active pharmacological agent bypasses the circulatory system and BBB, in addition to avoiding the risk of protein degradation that occurs after oral administration, intraperitoneal injection, or intravenous injection (Meredith et al., 2015). Thus, IL-10 was allowed to act directly on the CNS, to counteract pro-inflammatory signaling originating from both the periphery and the CNS while its main presence remained in the brain.

Because so little is known about the role IL-10 plays in the regulation of cognition or mood, this study sought to provide insight into whether IL-10 is pro-cognitive in a mouse model of depression, and if it provides antidepressant actions (for a working model of the hypotheses tested in this study, see Fig. 5). We developed a new mouse model that exhibits learning and memory impairments after induction of learned helplessness, which was associated with activated microglia and reduced dendritic spine density. These impairments were reversed by intranasal administration of IL-10, demonstrating the importance of IL-10 in regulating the brain's physiological response to stress.

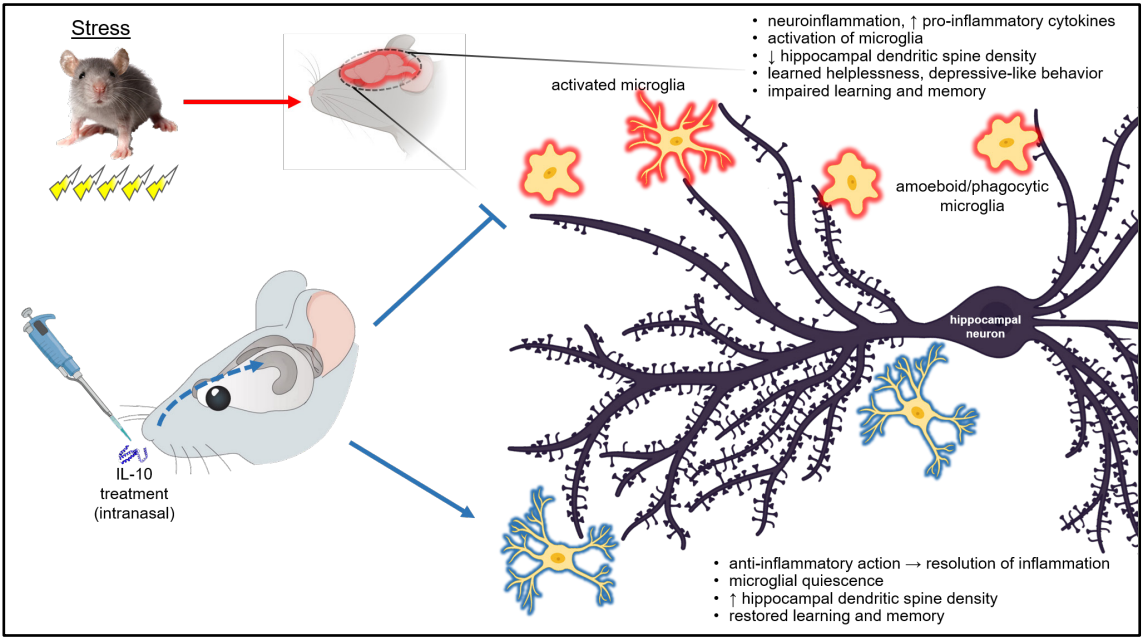

**Figure 5. Working hypothesis summarizing critical neurobiological consequences of stress and associated neuroinflammation for mood and cognition, as well as the theorized effect of IL-10 treatment**.

# Chapter 3 – Materials and Methods

## 3.1 Mice and drug administration

Studies investigating learning and memory impairments were conducted using male and female adult (8-12 weeks old) C57BL/6 wild-type mice in the context of stress while transgenic GSK3$\alpha$/$\beta^{21A/21A/9A/9A}$ knock-in and $FMR1^{-/-}$ mice were used as models of cognitive impairment in the absence of stress. Relevant data pertaining to female mice are found in Figure 16. When indicated, mice were treated intranasally (i.n.) with phosphate-buffered saline (PBS) or murine recombinant IL-10 (5 μg/mouse, Peprotech) 24 h and 1 h prior to behavioral testing. When indicated, mice received 10 μg/mouse siRNA targeting STAT3 (69367, Ambion) or scrambled siRNA control (Silencer Negative Control #5 siRNA, Ambion) 24 h prior to subjecting mice to behavioral testing and every 24 h for the subsequent 3 days until the end of the experiment. Depletion of microglia was achieved using the CSF1R inhibitor PLX5622 (1200 ppm, Plexxikon) present in the diet (AIN-76A). Mice were fed ad libitum for a minimum of 3 days before each experiment and the diet was maintained for the duration of the experiment. Mice were bred in the University of Miami animal facility. Mice were housed in standard cages in light and temperature-controlled rooms and were treated in accordance with NIH and the University of Miami Institutional Animal Care and Use Committee regulations.

## 3.2 Behaviors

### 3.2.1 Learned helplessness

The learned helplessness paradigm was used to induce depressive-like behavior in mice as described previously (Beurel et al., 2018). Mice were placed in one of two chambers (Med Associates, St. Albans, VT, USA) separated by a closed gate. Mice received 180 inescapable foot-shocks (IES) of 0.3 mA intensity and an average duration of 6-10 s with random and unpredictable inter-shock intervals ranging from 5 to 45 s. On the following day, mice were reintroduced to the same apparatus. Once the 0.3 mA foot-shocks began, the gate separating both chambers opened, giving each mouse a maximum of 24 s to escape the shock by shuttling to the opposite chamber. The shock was terminated when the mouse crossed the gate. Mice were tested in 30 escapable foot-shock (ES) trials and the mice that failed to escape more than 15 out of 30 trials were considered learned helpless.

### 3.2.2 Learning and memory

Mice were acclimated to the behavioral testing room for at least 30 min prior to testing in the presence of a white noise generator (55 dB) and were assessed twice for each behavior, before and after treatment. All sessions were video-recorded, and all videos were analyzed blind to the treatment.

#### 3.2.2.1 Novel object recognition (NOR)

A baseline measurement of NOR was conducted 1 day after the last escapable foot-shock trials. This task takes advantage of a mouse's natural inclination to preferentially explore novel objects it encounters compared to previously encountered, familiar objects. The amount of time mice spent exploring a novel object compared to a familiar object provides an indication of recognition memory (Antunes & Biala, 2012). Two identical copies of object 1 (125-mL glass bottles with black cap, 4-6 cm in diameter × 10 cm in height) were placed in a Plexiglas arena (50 cm long × 20 cm wide × 25 cm tall) equidistant from the walls of the container. Mice were individually placed within the arena and were allowed to explore the objects during a 5-min habituation phase. Following this, mice were removed from the arena and placed in an opaque holding container for 5 min. Mice were reintroduced to the same arena and presented with one copy of object 1 as well as a novel, never previously encountered object 2 (translucent red Plexiglas bottle, 8 cm in diameter × 13 cm in height). The total time each mouse spent exploring object 1 (familiar, F) compared to object 2 (novel, N) was recorded (exploration was defined as touching the object with the nose or forelimbs, sniffing the object, or closely approaching the object in a forward, attention-directed lunge) as we previously reported (Pardo et al., 2015). Both the percent of time spent with each object and the discrimination ratio [(time spent with N-time spent with F)/(time spent with N and F)] was reported. After IL-10 treatment, NOR was re-assessed with a new set of objects (object 1: blue bottle cap with metal nozzle, 6 cm in diameter × 4 cm in height; object 2: black rubber bottle stopper, 4 cm in diameter × 7 cm in height) in the same cohort of mice.

### 3.2.2.2 Two-trial Y-maze

A two-trial Y-maze was used to measure spatial working memory (Dudchenko, 2004). The Y-maze apparatus (Stoelting Co.) (lane width 5 cm × arm length 35 cm × arm height 10 cm) was arranged on an elevated platform with one arm facing outward toward the experimenter (start arm, S) and two distal arms facing inward toward the back wall of the room. Extra-maze visual cues of varying shapes and sizes were placed on the walls of the room ranging from 0.5 to 2 m distance from the maze itself. During a preliminary acquisition phase, the left or the right distal arm was obstructed by placing a barrier block at the entrance of the arm. Mice were individually placed into the end of the start arm and allowed to explore the open areas of the maze for 5 min. Following this, mice were removed from the maze and returned to their home cage during a 30-min inter-trial interval (ITI). During a 2-min retrieval trial, mice were returned to the start arm and allowed to freely explore all arms, including the previously obstructed, novel arm (N). Measurement of exploratory behavior began when a mouse had left the start arm and the total time each mouse spent exploring the novel arm compared to the familiar arm was calculated. An arm entry was defined as placing all four paws within the arm and a mouse was considered to have exited an arm when all four paws were located outside of the arm. Periods in which a mouse engaged in self-grooming or remained stationary were excluded from the final calculation. After IL-10 treatment, mice were reassessed in the Y-maze in a different room with a distinct set of extra-maze visual cues. Mice were assessed using a 1-min ITI between acquisition and retrieval trials to confirm preference for novelty, sufficient visual acuity to recognize extra-maze visual cues, and to control for potential anxiety-like motivational disturbances that may have influenced exploratory behavior.

### 3.3    Flow cytometry

After behavioral testing, mice were anesthetized in a chamber filled with an overdose of vaporized isoflurane until respiration ceased (within 2 min), were transcardially perfused with PBS, and hippocampi were recovered. The hippocampi were passed through a 70-μm cell strainer (BD Bioscience) and the cell suspension was mixed (vol/vol) to obtain a 30% Percoll/R1 medium that was overlaid on 70% Percoll/R1 medium in a centrifuge tube, and centrifuged at 2000 rpm for 20 min without using the brake. The cells at the interface of the 30/70% Percoll gradient were recovered, washed one time, resuspended in R10 media, aliquoted into a 96 well plate, and incubated for 4 h with the Protein Transport Inhibitor Cocktail (eBioscience) at the recommended concentrations. Standard intracellular cytokine staining was carried out as described (Harrington et al., 2005) using the Staining Intracellular Antigens for Flow Cytometry Protocol (eBioscience). Cells were first stained extracellularly with FITC-conjugated anti-CD45 (to identify microglial cells), and then were stained intracellularly with PcBlue-conjugated anti-IL-10. Samples were acquired on a Celesta (BD) flow cytometer and data were analyzed with the FlowJo software (Tree Star, Inc.).

### 3.4    Immunofluorescence

After subjecting mice to the learned helplessness paradigm, mice were anesthetized in a chamber filled with an overdose of vaporized isoflurane until respiration ceased (within 2 min), were transcardially perfused with 0.9% NaCl and 4% paraformaldehyde (PFA; #P614, Sigma-Aldrich). Brains were extracted and placed in 4% PFA overnight at 4°C and stored in 30% sucrose, 0.02% sodium azide in phosphate buffer (pH 7.4) until sectioning. Brains were rapidly embedded in optimal cutting temperature compound (O.C.T., Tissue-Tek) in a mold (Sigma-Aldrich). Frozen sagittal brain sections of 20 μm thickness were prepared using a cryostat (Leica). Free-floating sections were washed in PBS (pH 7.5), incubated in PBS containing 0.1% Triton-X for 10 min, and blocked in PBS containing 5% normal donkey serum for 1.5 h. Sections were then incubated overnight at 4°C in primary antibody of anti-AIF-1/Iba1 (1:500, #NB100-1028, Novus Biologicals) or anti-STAT3 (1:1000, #12640, Cell Signaling Technology). Sections were washed in PBS, incubated with donkey anti–goat Alexa 594-conjugated secondary antibody (1:1000; Life Technologies) or chicken anti-rabbit Alexa 488-conjugated secondary antibody (1:1000; A-21221 Life Technologies) overnight at 4°C. Following washes in PBS, tissue was mounted on a glass microscope slide and cover-slipped using VectaShield mounting medium with DAPI (Vector Laboratories). Hippocampal images were acquired using an Olympus    FLUOVIEW    FV1000    confocal    laser-scanning    microscope    with    40X magnification housed at the Microscopy Core Facility, or a VS120 Olympus slide scanner with 40X magnification housed at the Analytical Imaging Shared Resource, at the University of Miami. Individual confocal stacks were merged to form composite images containing the entire hippocampus, which were then analyzed blinded to treatment using standardized parameters with the ImageJ software. For each mouse, quantification of Iba1-positive microglia was performed by counting the total number of positive cells within a sampling area comprised of 5 separate 200 μm x 200 μm square ROIs (combined total area of 0.2 mm²) superimposed and distributed randomly over the dentate gyrus. Sampling areas within 2-4 hippocampal slices per animal were analyzed and the average number of Iba1-positive microglia within the dentate gyrus was determined for each mouse. Furthermore, microglial sphericity was evaluated in the dentate gyrus using Image J/Fiji (version 1.53). Ten well-resolved microglial cells were analyzed per image in 2-4 hippocampal slices per animal. The perimeter of the cell body was traced and microglial cross-sectional area ($\mu m^2$) was calculated as a measure of soma size. Additionally, the length (μm) of the longest cell process emanating from the cell body was measured to indicate the scale of cell process branching.

### 3.5    Golgi staining

After subjecting mice to the learned helplessness paradigm, mice were sacrificed and brains were recovered to undergo Golgi staining according to the manufacturer's instructions (FD Neurotechnologies, Inc.). Once fixed, brains were flash-frozen and 120-μm coronal sections were prepared and mounted onto gelatin-coated glass slides. After allowing the tissue to dry and adhere to the slides for 1 week at 35°C, samples were dehydrated with sequentially increasing percentage ethanol/ddH$_2$O solutions and xylene, and finally slides were cover-slipped using Eukitt (#NC9068612, Thermo Fisher Scientific) hardening mounting medium. Hippocampal images were acquired at 100X magnification with the EvosXL Core light microscope (Life Technologies). At Bregma -2.30 mm, apical dendrites, 20 µm from the neuronal cell body, were imaged within the CA1, CA3, and dentate gyrus anatomical regions of the dorsal hippocampus. Images were analyzed blinded to treatment and dendritic spine density was quantified using the Neurolucida (version 11, MBF Bioscience) and Neuroexplorer software or ImageJ/Fiji (version 1.53). The average number of dendritic spines over approximately 80 µm of total apical dendrite length (not necessarily from the same neuron) per hippocampal region was determined for each mouse.

### 3.6    Immunoblot analysis

Mice were sacrificed by decapitation 1 h after i.n. administration of either vehicle (PBS) or IL-10 (5 µg/mouse). Brains were extracted and prefrontal cortex, hippocampus, and cerebellum were dissected. Samples were homogenized in lysis buffer containing 20 mM Tris–HCl, pH 7.4, 150 mM NaCl, 1 mM EDTA, 1 mM EGTA, 1% Triton-100, 10 µg/mL leupeptin, 10 µg/mL aprotinin, 5 µg/mL pepstatin, 1 mM phenylmethanesulfonylfluoride fluoride, 1 mM sodium orthovanadate, 50 mM sodium fluoride, and 100 nM okadaic acid. Tissue homogenates from the 3 brain regions were centrifuged at 14,000 rpm for 10 min at 4°C to remove cellular debris and the supernatant was recovered. Protein concentration was evaluated using the Bradford method. Thirty micrograms of protein was loaded onto a 10% SDS polyacrylamide gel and transferred into a nitrocellulose membrane. The membrane was blocked and incubated with primary anti-phospho-Tyr[705]-STAT3 or anti-STAT3 antibodies overnight at 4°C, followed by an overnight incubation at 4°C with a secondary antibody conjugated to horseradish peroxidase (HRP). Signals were visualized using chemiluminescence with an Amersham Imager600 and quantified using iQTL software (GE Healthcare). Blots were reblotted for β-actin to ensure proper loading.

### 3.7    Statistical analysis

Data were represented as mean ± SEM. Results were analyzed using a Student's *t*-test or one- or two-way analysis of variance (ANOVA) with a Bonferroni or Fisher's LSD *post-hoc* test using the Prism software as indicated; p-values of less than 0.05 were considered statistically significant (Table 3).

# Chapter 4 – Results

## 4.1  Microglia were activated after learned helplessness

Because stress induces microglial activation (Wohleb et al., 2018), we first tested if microglial cells were activated after the induction of learned helplessness. Exposure to the learned helplessness protocol produced two groups of mice, those that exhibited learned helplessness (mice failed to escape >15/30 trials, LH mice), and mice that did not exhibit learned helplessness (mice failed to escape <15/30 trials, NLH mice) (Fig. 6A). These two groups of mice were compared to non-shocked (NS) mice following Iba1 immunostaining to detect activated microglial cells. The number of Iba1$^+$ microglial cells in the dentate gyrus of the hippocampus increased ~1.7 to 2-fold 48 h after the learned helplessness protocol in both NLH and LH mice compared to NS mice (Fig. 6C-F). This indicates that foot-shocks are sufficient to increase the number of Iba1$^+$ microglial cells. Similar increases in the number of Iba1$^+$ microglial cells were observed in the CA1 and CA3 regions of the hippocampus (data not shown) suggesting that microglia were activated by foot-shocks throughout the hippocampus. In addition, Iba1$^+$ microglial cells in the dentate gyrus displayed an amoeboid morphology as exemplified by a bigger soma area, in mice subjected to the learned helplessness paradigm, whether they exhibited learned helplessness or not, compared to non-shocked mice that had microglia that exhibited a more quiescent, ramified phenotype (Fig. 6C-E, G). These results indicate that stress caused by foot-shocks activates hippocampal microglial cells, which may contribute to behavioral responses to stress.

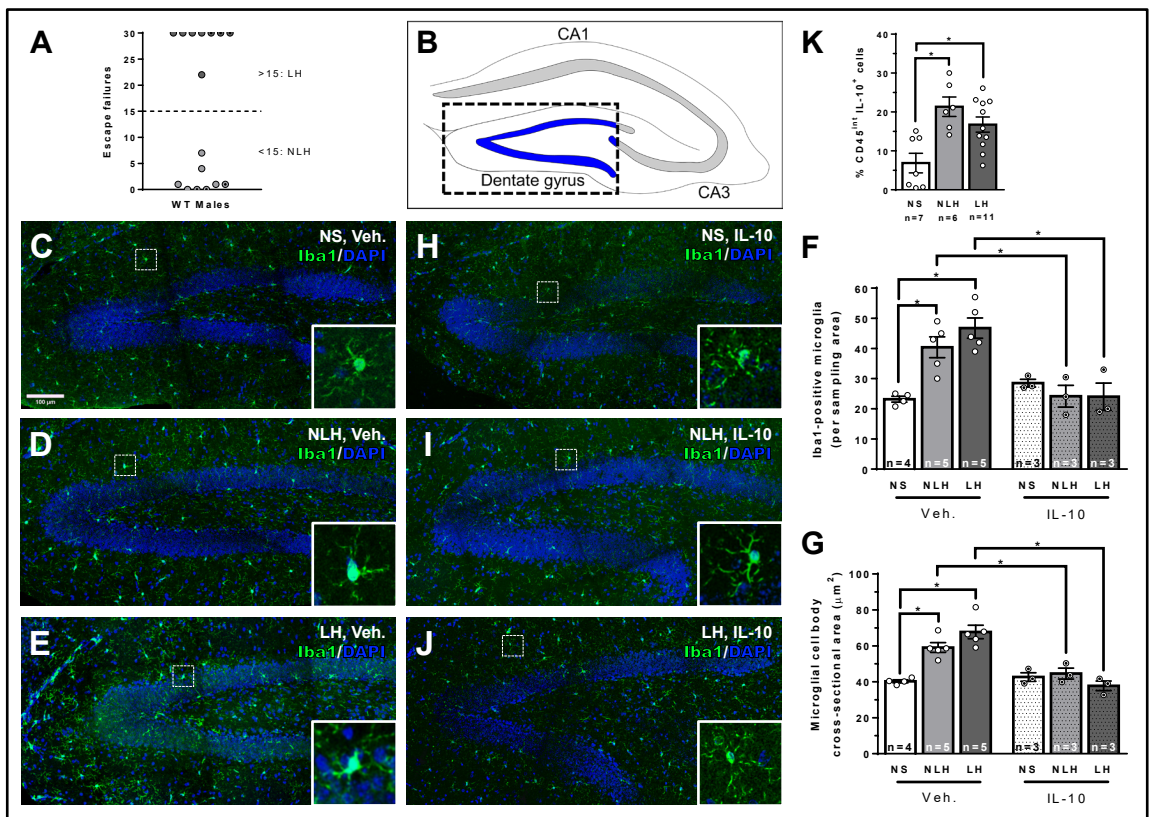

**Figure 6. Microglial activation increased in stressed mice and was alleviated by IL-10 treatment.** Male wild-type mice were subjected or not (NS) to the learned helplessness paradigm and separated into 2 groups: non-learned helpless (NLH) and learned helpless (LH) mice, according to their number of failures out of 30 escapable shock trials **(A)** and were treated or not with vehicle (Veh.) or IL-10 (5 μg/mouse) 24 h and 1 h prior to sacrifice. Mice were perfused and brains were collected for immunostaining. **B** Diagram of the dorsal mouse hippocampus in cross-section highlighting three major anatomical regions of the perforant pathway: CA1, CA3, and the dentate gyrus (DG). Representative images of Iba1[+] microglial cells in the dentate gyrus of vehicle-treated: non-shocked control (NS) **(C)**, NLH **(D)**, and LH **(E)** mice, and IL-10-treated: NS **(H)**, NLH **(I)**, and LH **(J)** mice. Enlarged images of representative microglial cells (white dashed outline) exhibiting the predominant morphology are included in each panel. **F** Number of Iba1[+] microglial cells per 200,000 μm² sampling area within the dentate gyrus for each mouse. Each dot represents a mouse. Two-way ANOVA, $F_{(2,17)interaction}=9.498$, $F_{(1,17)treatment}=17.09$, $F_{(2,17)condition}=4.181$, Bonferroni post-hoc test, *$p<0.05$, bars represent mean ± SEM, $n=3-5$ mice/group. **G** Microglial morphology (sphericity) was measured as the microglial cell body cross-sectional area within the dentate gyrus for each mouse. Each dot represents a mouse. Two-way ANOVA $F_{(2,17)interaction}=13.99$, $F_{(1,17)treatment}=32.02$, $F_{(2,17)condition}=8.281$, Bonferroni post-hoc test, *$p<0.05$, bars represent mean ± SEM, $n=3-5$ mice/group. **K** Percentage of CD45[int]IL-10[+] microglial cells in the hippocampus of NS, NLH, and LH mice, measured by flow cytometry. Each dot represents a mouse. One-way ANOVA, $F_{(2,21)}=8.895$, Bonferroni post-hoc test, *$p<0.05$, bars represent mean ± SEM, $n=6-11$ mice/group.

## 4.2    IL-10 production increased in stressed mice

We also tested if activated microglia produce anti-inflammatory IL-10 after learned helplessness, using flow cytometry to identify IL-10[+] microglial cells (Fig. 7A). We found that hippocampal IL-10[+]CD45int cells were significantly 3- and 2.4-fold increased in NLH and LH mice respectively compared to NS mice (Fig. 6K). The proportion of microglial cells expressing IL-10 was reduced by 20% in the LH mice compared to the NLH mice. This is consistent with previous reports of lower brain IL-10 in mice that display depressive-like behavior and demonstrate that deficient microglial IL-10 contributes to this deficit (Mesquita et al., 2008).

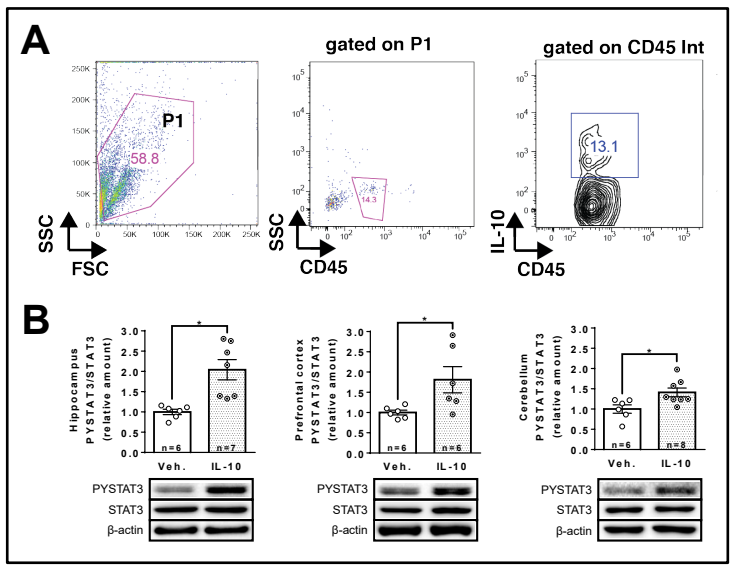

**Figure 7. IL-10 promoted activation of STAT3 in the hippocampus, cerebral prefrontal cortex, and cerebellum. (A)** Gating strategy used for the analysis of IL-10+CD45Int cells by flow cytometry after learned helplessness. **(B)** Male mice were treated intranasally with either vehicle (Veh.) or IL-10 (5 μg/mouse) for 1 h. Phospho-Tyr705-STAT3 (PYSTAT3) and STAT3 proteins were immunoblotted in the hippocampus, cerebral prefrontal cortex, and cerebellum. Membranes were reblotted for b-actin to ensure proper loading. Quantification of the samples on the top were represented as the ratio of PYSTAT3/STAT3 in each brain region, and a representative image of the western blot was shown in the bottom. Mann-Whitney test, U=0, 4, and 3, *p<0.05 compared to vehicle-treated mice, bars represent mean ± SEM, n=6-8.

### 4.3    IL-10 prevented foot-shock-induced microglial activation

To ensure microglial activation was associated with a microglial inflammatory profile, we tested if administration of the anti-inflammatory cytokine IL-10 was capable of reversing the activated phenotype of microglial cells in the dentate gyrus. We administered IL-10 intranasally and found IL-10-downstream signaling, STAT3 Tyr705-phosphorylation increased 1 h after IL-10 treatment in several brain regions, including the hippocampus (Fig. 7B), indicating that intranasal administration of IL-10 lead to STAT3 activation in the brain. Intranasal IL-10 treatment did not affect Iba1 staining in NS mice, but reduced microgliosis by 40% and 48% in the dentate gyrus of NLH and LH mice, respectively, compared to vehicle-treated NLH and LH mice (Fig. 6F, H-J), reaching similar levels of Iba1$^+$ cells detected in NS mice. These results indicate that stress promotes microglial activation that can be alleviated with IL-10 treatment.

**Figure 8. The decrease of hippocampal DG dendritic spine density in stressed mice was abolished by IL-10 treatment.** Male wild-type mice were subjected or not (NS) to the learned helplessness paradigm and separated into 2 groups: non-learned helpless (NLH) and learned helpless (LH) mice, according to their number of failures out of 30 escapable shock trials and were treated or not with vehicle (Veh.) or IL-10 (5 µg/mouse) 24 h and 1 h prior to sacrifice. **(A)** Representative images of dendritic spines on apical dendrites radiating from granule cells within the dentate gyrus (DG). **(B)** Dendritic spine density was calculated as an average per animal (# spines/µm) over 80 µm of total analyzed apical dendrite length (20 µm from the cell body) extending from granule cells within the molecular layer of the DG region of the dorsal hippocampus. Each dot represents a mouse. Two-way ANOVA, $F_{(2,21)interaction}$ = 13.33 , $F_{(1,21)treatment}$ = 47.51, $F_{(2,21)condition}$ = 6.920, Bonferroni post-hoc test, *$p<0.05$, bars represent mean ± SEM, $n$=5-8 mice/group.

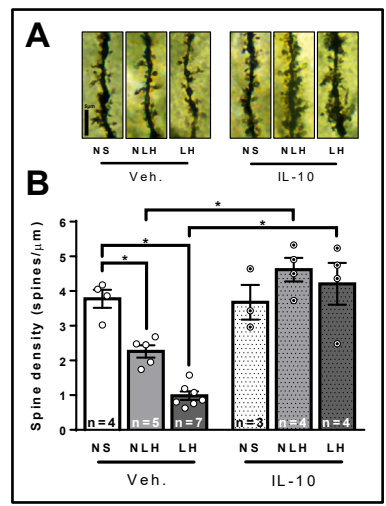

## 4.4 Dendritic spine density was reduced in learned helpless mice, and IL-10 administration promoted dendritic spine density

Because microglia are important for synaptic pruning (Paolicelli et al., 2011), we examined if dendritic spine density was affected by the induction of learned helplessness, which activated microglial cells. Consistent with the findings that stress reduces dendritic spine density in the hippocampus (Shirayama et al., 2002), we found that dendritic spine density was decreased by 1.8-fold in the dentate gyrus of NLH mice, and was decreased significantly more, 3.2-fold, in the dentate gyrus of LH mice (Fig. 8). Similar differences were evident in the CA1 and CA3 regions of the hippocampus (Fig. 9). These findings confirm that foot-shock stress is sufficient to reduce hippocampal dendritic spine density, and further demonstrate that greater hippocampal dendritic spine density reductions are evident in mice that develop learned helplessness.

Intranasal administration of IL-10 increased dendritic spine density by 2.0- and 4.3-fold in the dentate gyrus of NLH and LH mice, respectively, compared to vehicle administration. Furthermore, after IL-10 treatment the dendritic spine density was increased 1.3- and 1.1-fold in the dentate gyrus of NLH and LH mice compared to NS mice treated with IL-10 (Fig. 8), suggesting that IL-10 enhanced dendritic spine density in the hippocampus to at or above basal levels. Altogether, this data suggests that since hippocampal dendritic spine density was greatly reduced in LH mice, cognitive performance after learned helplessness might also be impaired, and that IL-10 administration might reverse such impairments.

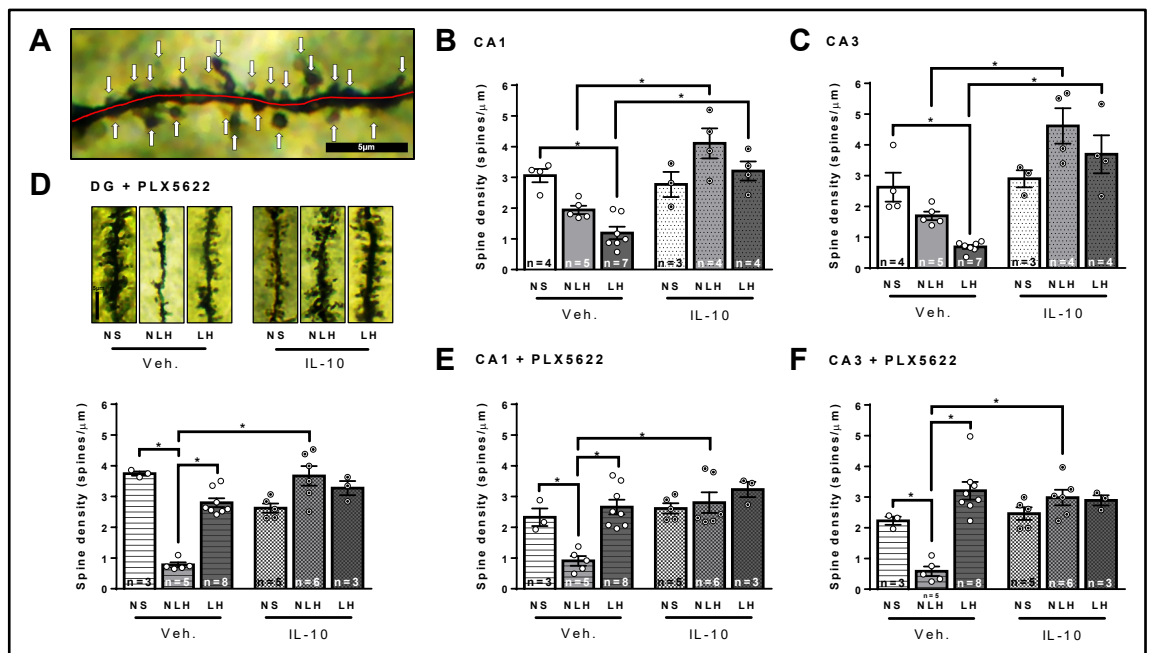

**Figure 9. The decrease of hippocampal CA1 and CA3 dendritic spine density in stressed mice was abolished by IL-10 treatment.** Male wild-type mice were subjected or not (NS) to the learned helplessness paradigm and separated into 2 groups: non-learned helpless (NLH) and learned helpless (LH) mice, according to their number of failures out of 30 escapable shock trials and were treated or not with vehicle (Veh.) or IL-10 (5 µg/mouse) 24 h and 1 h prior to sacrifice. **(A)** Representative image of dendritic spines on apical dendrites radiating from granule cells within the dentate gyrus (DG) of a NLH mouse. Arrows point to the actual dendritic spines that were recorded. **(B)** Dendritic spine density was calculated as an average per animal (# spines/µm) over 80 µm of total analyzed apical dendrite length (20 µm from the cell body) extending from granule cells within the molecular layer of the dentate gyrus and pyramidal cells within the stratum radiatum of the CA1 **(B)** and CA3 **(C)** regions of the dorsal hippocampus. Each dot represents a mouse. Two-way ANOVA $F_{(2,21)interaction}=9.738$, $F_{(1,21)treatment}=29.09$, $F_{(2,21)condition}=5.171$, Bonferroni post-hoc test, *$p<0.05$, bars represent mean ± SEM, $n=5-7$ mice/group. Male mice were fed with the AIN-76A diet supplemented with PLX5622 for 3 days before the induction of the learned helplessness paradigm and were maintained on the diet for the duration of the experiment. 3 days after initiating the diet, mice were subjected or not (NS) to the learned helplessness paradigm and separated into 2 groups: non-learned helpless (NLH) and learned helpless (LH) mice, according to their number of failures out of 30 escapable shock trials, and the next day learning and memory was assessed. After the cognitive assessments, mice were treated with IL-10 (5 µg/mouse) i.n. and retreated 1 h before the learning and memory re-assessment on the next day as shown in Fig. 11. Dendritic spine densities in **(D)** dentate gyrus (plus representative images), **(E)** CA1, and **(F)** CA3 were calculated. Each dot represents a mouse. Two-way ANOVA, $F_{(2,24)interaction}=44.31$, $F_{(1,24)treatment}=17.43$, $F_{(2,24)condition}=12.09$, Bonferroni post-hoc test, *$p<0.05$, bars represent mean ± SEM, $n=3-8$ mice/group.

## 4.5 Learned helpless mice exhibited microglia-dependent impaired learning and memory

We tested if exposure to the learned helplessness protocol affected learning and memory in two tasks, novel object recognition and spatial working memory (Fig. 10A).

We were particularly interested in determining if cognitive impairments were linked to the development of learned helplessness or were merely caused by the stress of the protocol. Mice were subjected to the learned helplessness paradigm and 24 h after the last escapable foot-shocks, novel object recognition or spatial working memory were tested in two different cohorts of mice. LH mice exhibited impaired novel object recognition, as they spent similar amounts of time exploring the familiar and the novel objects and they exhibited a reduced discrimination index (Fig. 10B and Fig. 11A). In contrast, NLH mice exhibited preference for the novel object equivalent to NS mice, demonstrating intact novel object recognition. All mice spent similar amounts of time exploring the objects (Fig. 10C). These findings indicate that the stress of the LH protocol did not impair novel object recognition, but impairment was evident in mice that displayed learned helplessness.

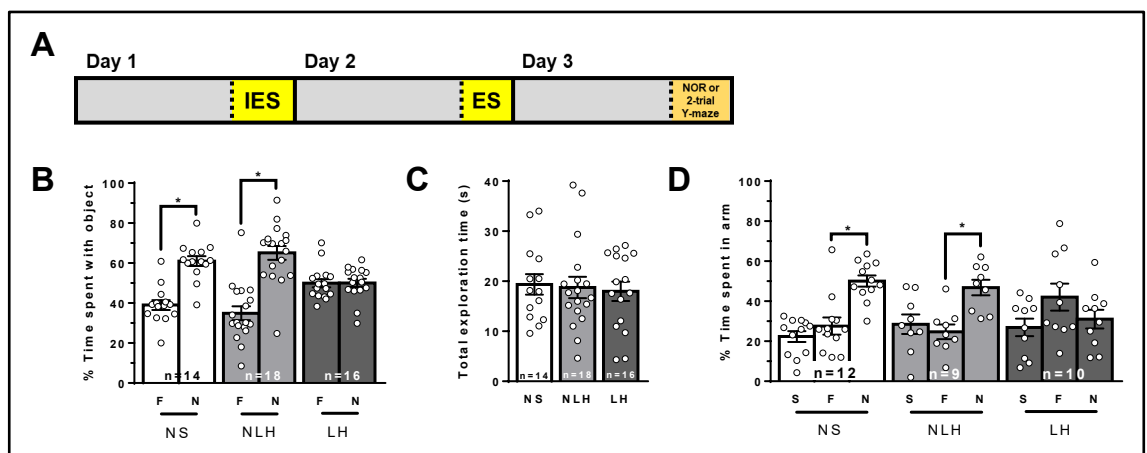

**Figure 10. Learned helpless mice exhibited impaired novel object recognition and spatial memory.** Male wild-type mice were subjected or not (NS) to the learned helplessness paradigm and separated into 2 groups: non-learned helpless (NLH) and learned helpless (LH) mice, according to their number of failures out of 30 escapable shock trials, and learning and memory were assessed the next day. **(A)** Timeline of the experiment. The percentage of a mouse's total object exploration time spent exploring the familiar (F) and novel (N) objects **(B)** and the total time (s) mice spent engaged in exploration of both the familiar and novel objects combined **(C)** in the novel object recognition test were reported. Each dot represents a mouse. Two-way ANOVA, $F_{(2,90)interaction}=15.53$, $F_{(1,90)treatment}=56.16$, $F_{(2,90)condition}=4.238\times10^{-14}$, Bonferroni post-hoc test, $*p<0.05$, compared to % time spent with familiar object, bars represent mean ± SEM, $n=14-18$ mice/group. **(D)** The percentage of a mouse's total maze exploration time spent exploring the start (S), familiar (F), and novel (N) arms in the two-trial Y-maze test was measured in a different cohort of mice than B-C. Each dot represents a mouse. Two-way ANOVA, $F_{(4,84)interaction}=5.373$, $F_{(2,84)treatment}=11.68$, $F_{(2,84)condition}=7.654\times10^{-14}$, Fisher's least significant difference test, $*p<0.05$, compared to % time spent exploring familiar arm, bars represent mean ± SEM, $n=9-12$ mice/group.

**Figure 11. The impairment seen in learned helpless mice was not the result of anxiety or inability to recognize the cues. (A)** Male mice were fed or not with the AIN-76A diet (control diet), the AIN-76A diet supplemented with PLX5622 for 3 days before the induction of the learned helplessness paradigm and were maintained on the diet for the duration of the experiment. 3 days after initiating the diet, mice were subjected or not (NS) to the learned helplessness paradigm and separated into 2 groups: non-learned helpless (NLH) and learned helpless (LH) mice, according to their number of failures out of 30 escapable shock trials, and the next day learning and memory was assessed. After the cognitive assessments, mice were treated with IL-10 (5 μg/mouse) i.n. and retreated 1 h before the NOR assessment on the next day. The discrimination index is represented. Each dot represents a mouse. Two-way ANOVA, F(2,82)$_{interaction}$= 4.6, F(1,82)$_{treatment}$= 2.9, F(2,82)$_{condition}$=3.0, Bonferroni post-hoc test, *$p$<0.05, bars represent mean ± SEM, $n$=11-18 mice/group; Two-way ANOVA, F(4,57)$_{interaction}$= 4.8, F(2,57)$_{treatment}$= 2.0, F(2,57)$_{condition}$= 3.04, Bonferroni post-hoc test, *$p$<0.05, bars represent mean ± SEM, $n$=4-12 mice/group. Male mice were fed or not with the AIN-76A diet supplemented with PLX5622 for 3 days before the induction of the learned helplessness paradigm and were maintained on

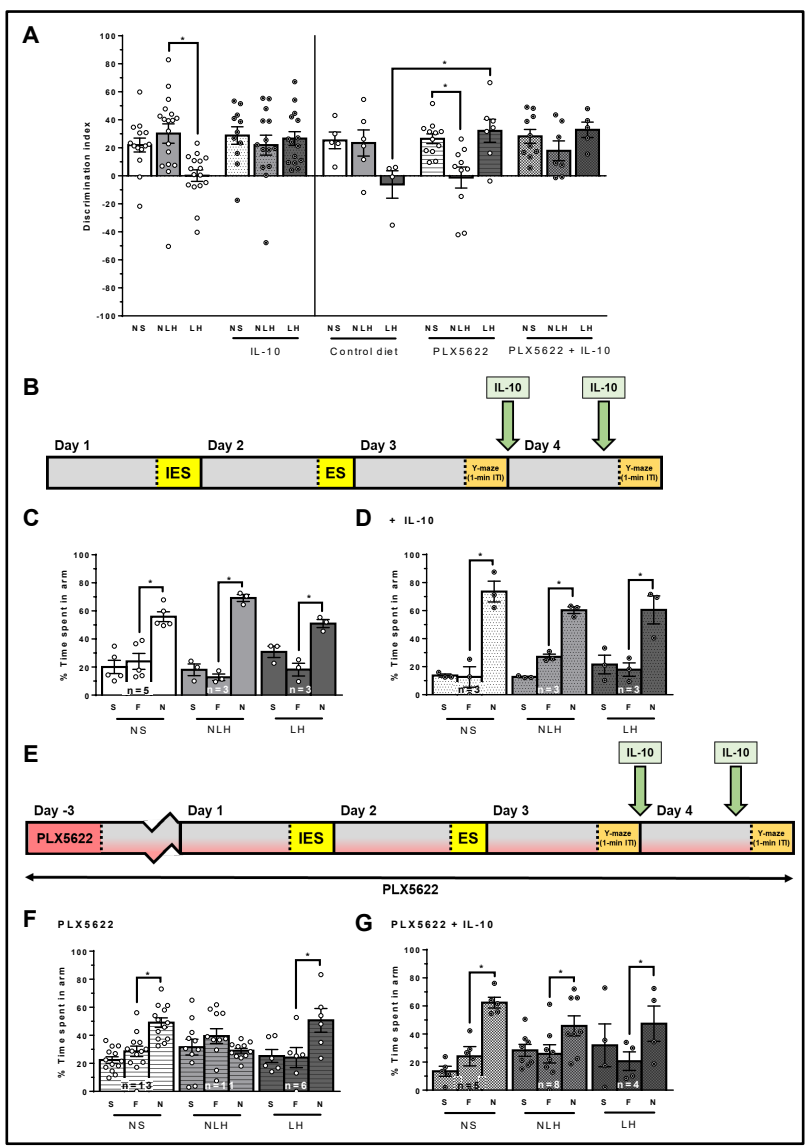

the diet for the duration of the experiment. 3 days after initiating the diet, mice were subjected or not (NS) to the learned helplessness paradigm and separated into 2 groups: non-learned helpless (NLH) and learned helpless (LH) mice, according to their number of failures out of 30 escapable shock trials, and the next day learning and memory was assessed as described in **B** and **E**. After the first cognitive assessment (C), mice were treated with IL-10 (5 μg/mouse) i.n. and retreated 1 h before the second spatial memory re-assessment on the next day. The percentage of a mouse's total maze exploration time spent exploring the start (S), familiar (F), and novel (N) arms using a modified 1-min inter-trial interval (ITI) in the two-trial Y-maze test was measured in different cohorts of mice receiving **(F-G)** or not **(C-D)** PLX5622. Each dot represents a mouse. Two-way ANOVA F(4,24)$_{interaction}$=3.685, F(2,24)$_{treatment}$=68.48, F(2,24)$_{condition}$= 8.47×10$^{-11}$ **(C)**, F(4,18)$_{interaction}$= 2.102, F(2,18)$_{treatment}$= 70.25, F(2,18)$_{condition}$= 1.901×10$^{-13}$ **(D)**, F(4,27)$_{interaction}$= 2.219, F(2,27)$_{treatment}$= 107.1, F(2,27)$_{condition}$= 7.569×10$^{-11}$ **(F)**, F(4,21)$_{interaction}$= 1.188, F(2,21)$_{treatment}$= 58.12, F(2,21)$_{condition}$= 8.755×10$^{-14}$ **(G)**, Fisher's least significant difference test, *$p$<0.05, bars represent mean ± SEM, $n$=3-5 mice/group.

Similar results were obtained using the two-trial Y-maze, LH mice spent similar amounts of time in the familiar and novel arms, indicative of cognitive impairment. In contrast, both NLH and NS mice spent more time in the novel arm than the familiar arm (Fig. 10D). Yet, LH mice preferred the novel arm in the 1-min ITI protocol, suggesting that the impairment seen in LH mice was not the result of anxiety or inability to recognize the cues (Fig. 11). As with the novel object recognition task, this indicates that the development of learned helplessness and depressive-like behavior, but not exposure to the foot-shock protocol, caused learning and memory impairment. Thus, LH mice exhibited impaired novel object recognition and impaired spatial memory that were not impaired in NLH mice exposed to the learned helplessness protocol.

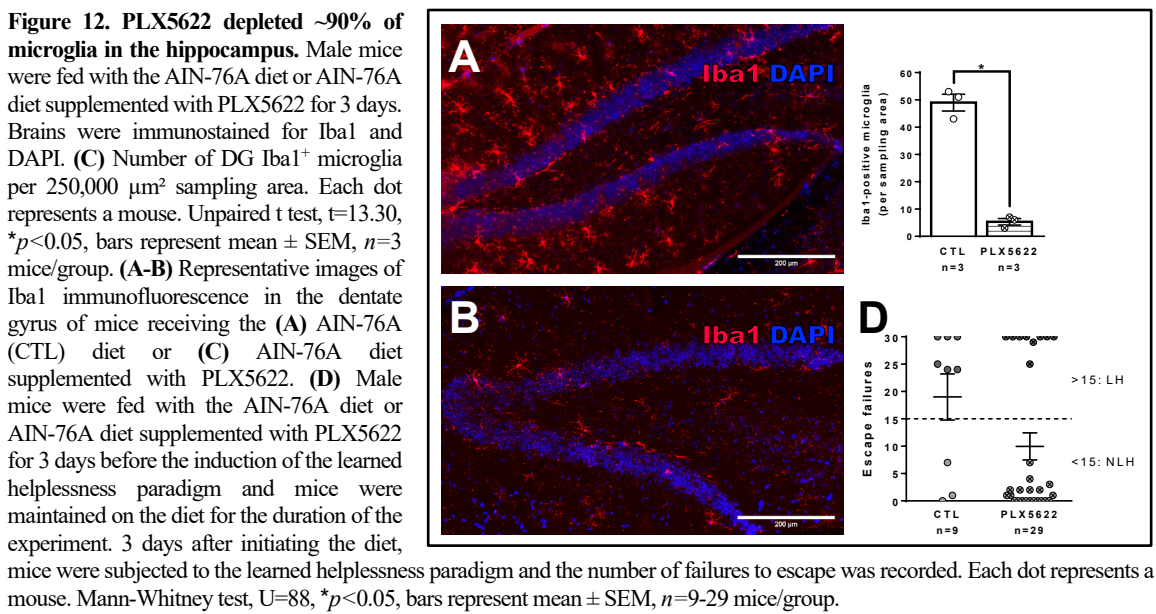

**Figure 12. PLX5622 depleted ~90% of microglia in the hippocampus.** Male mice were fed with the AIN-76A diet or AIN-76A diet supplemented with PLX5622 for 3 days. Brains were immunostained for Iba1 and DAPI. **(C)** Number of DG Iba1$^+$ microglia per 250,000 μm² sampling area. Each dot represents a mouse. Unpaired t test, t=13.30, *$p$<0.05, bars represent mean ± SEM, $n$=3 mice/group. **(A-B)** Representative images of Iba1 immunofluorescence in the dentate gyrus of mice receiving the **(A)** AIN-76A (CTL) diet or **(C)** AIN-76A diet supplemented with PLX5622. **(D)** Male mice were fed with the AIN-76A diet or AIN-76A diet supplemented with PLX5622 for 3 days before the induction of the learned helplessness paradigm and mice were maintained on the diet for the duration of the experiment. 3 days after initiating the diet, mice were subjected to the learned helplessness paradigm and the number of failures to escape was recorded. Each dot represents a mouse. Mann-Whitney test, U=88, *$p$<0.05, bars represent mean ± SEM, $n$=9-29 mice/group.

To test if microglial cells were involved in the cognitive impairments of LH mice, mice were treated for 3 days with PLX5622, which eliminated ~90% of microglial cells [Fig. 12A-C, (Acharya et al., 2016)], prior to behaviors (Fig. 13A), and similar elimination of microglia cells was found after 6 days of PLX5622 (data not shown). LH mice receiving the control diet (AIN-76A diet) without PLX5622 exhibited similar novel object recognition or spatial memory impairments as mice receiving the regular diet used by the animal facility at the University of Miami (Fig. 13B-D). Elimination of microglial cells with PLX5622 treatment did not affect novel object recognition or spatial memory of NS mice (Fig. 13E-G, Fig. 11A). However, PLX5622 treatment restored novel object recognition and spatial memory of LH mice (Fig. 13E-G, Fig. 11A), indicating that microglial cells contributed to these cognitive deficits associated with LH. PLX5622 also reduced the proportion of mice exhibiting learned helplessness by roughly half, from 66% in mice fed the control diet to 31% in PLX5622-treated mice (Fig. 12D). In line with evidence that microglia have important roles in healthy brain function as well as detrimental actions in diseased brains, microglial depletion with PLX5622 treatment impaired novel object recognition and spatial memory of NLH mice (Fig. 13E-G, Fig. 11A), which were not impaired in the absence of PLX5622 (Fig. 13B-D, Fig. 11A). These findings indicate that microglial cells in the LH mice contribute to impaired learning and memory, whereas microglial cells contribute to a healthy stress-response in supporting cognition in the stressed but NLH mice.

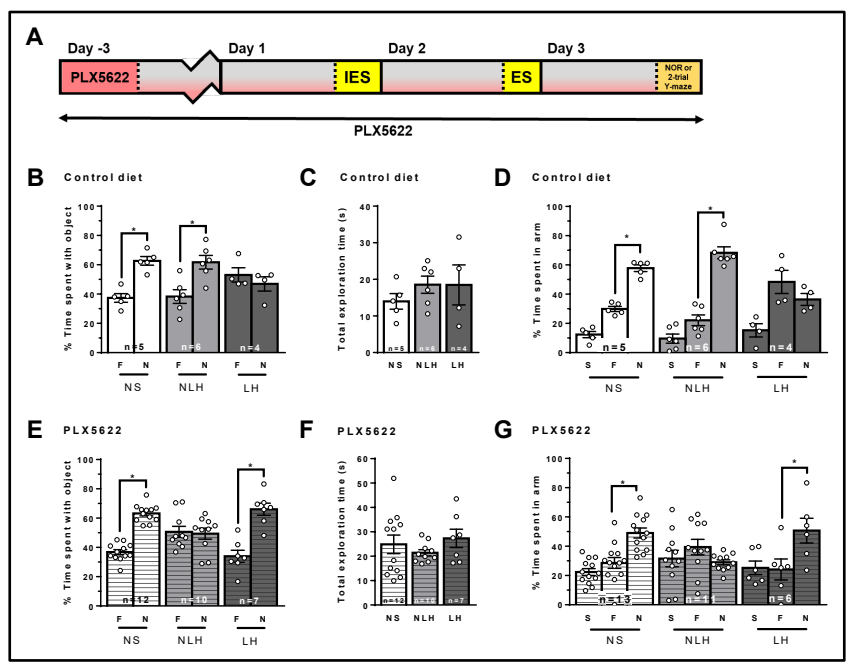

**Figure 13. Depletion of microglial cells was sufficient to rescue the learning and memory impairments of learned helpless mice.** Male mice were fed with the AIN-76A diet (control diet) **(B-D)** or AIN-76A diet supplemented with PLX5622 **(E-G)** for 3 days before the induction of the learned helplessness paradigm and were maintained on the diet for the duration of the experiment. Three days after initiating the diet, mice were subjected or not (NS) to the learned helplessness paradigm and separated into 2 groups: non-learned helpless (NLH) and learned helpless (LH) mice, according to their number of failures out of 30 escapable shock trials, and the next day learning and memory was assessed. **(A)** Timeline of the experiment. The percentage of a mouse's total object exploration time spent exploring the familiar (F) and novel (N) objects **(B, E)** and the total time (s) mice spent engaged in exploration of both the familiar and novel objects combined **(C, F)** in the novel object recognition test were reported. Each dot represents a mouse. Two-way ANOVA, $F(2,24)_{interaction}=7.334$, $F(1,24)_{treatment}=15.83$, $F(2,24)_{condition}=5.380\times10^{-14}$, Bonferroni post-hoc test, *$p<0.05$, compared to % time spent with familiar object, bars represent mean ± SEM, $n=4$-12 mice/group. **D, G** The percentage of a mouse's total maze exploration time spent exploring the start (S), familiar (F), and novel (N) arms in the two-trial Y-maze test was measured in a different cohort of mice than B and E. Each dot represents a mouse. Two-way ANOVA, $F(4,81)_{interaction}=5.807$, $F(2,81)_{treatment}=9.980$, $F(2,81)_{condition}=0.0$, Fisher's least significant difference test, *$p<0.05$, bars represent mean ± SEM, $n=4$-13 mice/group.

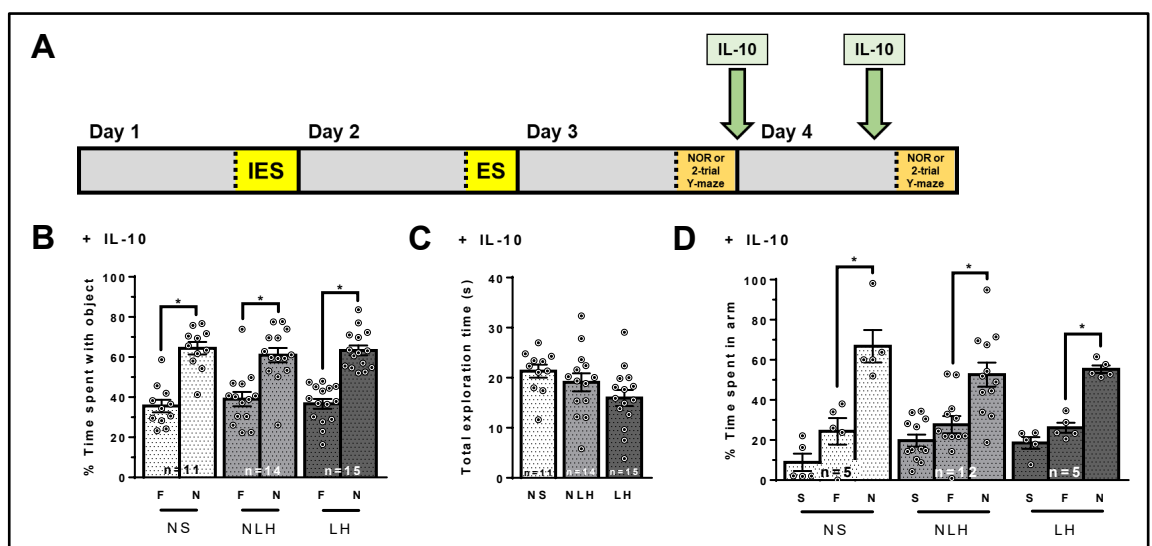

**Figure 14. IL-10 reversed learned helplessness-dependent recognition and spatial memory impairments.** Male mice were subjected or not (NS) to the learned helplessness paradigm and separated into 2 groups: non-learned helpless (NLH) and learned helpless (LH) mice, according to their number of failures out of 30 escapable shock trials, and the next day learning and memory was assessed. After the cognitive assessments, mice were treated with IL-10 (5 μg/mouse) i.n. and retreated 1 h before the learning and memory assessment on the next day as shown in **(A)**. The percentage of a mouse's total object exploration time spent exploring the familiar (F) and novel (N) objects **(B)** and the total time (s) mice spent engaged in exploration of both the familiar and novel objects combined **(C)** in the novel object recognition test were reported. Each dot represents a mouse. Two-way ANOVA $F_{(2,74)\text{interaction}}=0.6239$, $F_{(1,74)\text{treatment}}=102.8$, $F_{(2,74)\text{condition}}=4.298\times10^{-14}$, Bonferroni post-hoc test, $*p<0.05$, compared to % time spent with familiar object, bars represent mean ± SEM, $n=11$-15 mice/group. **(D)** The percentage of a mouse's total maze exploration time spent exploring the start (S), familiar (F), and novel (N) arms in the two-trial Y-maze test was measured in a different cohort of mice than B. Each dot represents a mouse. Two-way ANOVA, $F_{(4,57)\text{interaction}}=1.501$, $F_{(2,57)\text{treatment}}=46.50$, $F_{(2,57)\text{condition}}=4.036\times10^{-11}$, Fisher's least significant difference test, $*p<0.05$, bars represent mean ± SEM, $n=5$-12 mice/group.

## 4.6 IL-10 reversed learned helplessness-dependent recognition and spatial memory impairments

Since in LH mice microglial activation was reduced by treatment with IL-10 (Fig. 6F, H-J) and microglial cells contributed to the impaired learning and memory in LH mice (Fig. 10B-D), we tested if IL-10 administration reverses cognitive impairments in LH mice. Mice were subjected or not to the learned helplessness paradigm and treated with intranasal IL-10, 24 h and 1 h before cognitive assessment (Fig. 14A). Administration of IL-10 blocked the impairments in novel object recognition and spatial memory in LH mice, whereas IL-10 treatment did not affect novel object recognition or spatial memory of NS and NLH mice (Fig. 14B-D, Fig. 11A). IL-10 did not seem to affect learned helpless behavior per se as the percent of learned helpless mice after IL-10 treatment was similar to the one of mice after vehicle treatment (Fig. 15A).

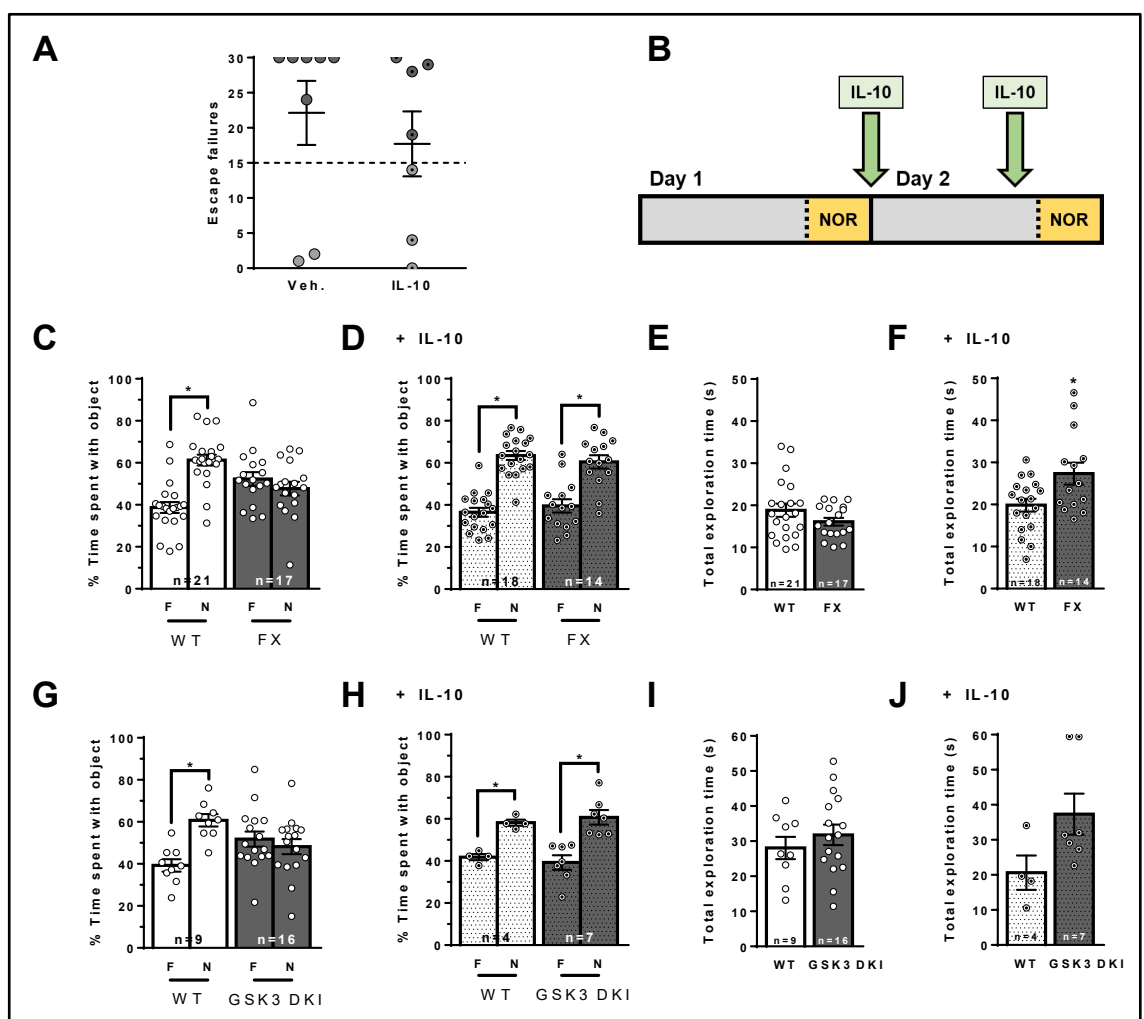

**Figure 15. IL-10 rescued the novel object recognition deficit of *FMR1*⁻/⁻ and GSK3α/β²¹ᴬ/²¹ᴬ/⁹ᴬ/⁹ᴬ knock-in mice.** (A) Male wild-type mice were subjected to the learned helplessness paradigm and separated into 2 groups: non-learned helpless (NLH) and learned helpless (LH) mice, according to the number of failures out of 30 escapable shock trials and were treated with IL-10 (5 μg/mouse) 24 h and 1 h prior to retesting the number of failures. Mann-Whitney test, U=17.50, *$p<0.05$, bars represent mean ± SEM, $n$=7-8 mice/group. Novel object recognition was assessed in male littermate wild-type (WT) and *FMR1*⁻/⁻ (FX) mice or GSK3α/β²¹ᴬ/²¹ᴬ/⁹ᴬ/⁹ᴬ knock-in (GSK3 DKI) mice, and treated intranasally with IL-10 (5 μg/mouse) after the first behavioral assessment and 1 h before the second behavioral re-assessment as shown in **(B)**. The percentage of a mouse's total object exploration time spent exploring the familiar (F) and novel (N) objects **(C-D, G-H)** and the total time (s) mice spent engaged in exploration of both the familiar and novel objects combined **(E-F, I-J)** in the novel object recognition task was reported for the FX mice **(C-F)** or the GSK3 DKI mice **(G-J)**. Each dot represents a mouse. One-way ANOVA, F(3,72)=11.56 **(C)**, F(3,60)=29.64 **(D)**, F(3,46)=4.436 **(G)**, F(3,18)=12.42 **(H)**, Bonferroni post-hoc test, *$p<0.05$, compared to % time spent with familiar object, Student's t test t=2.629 **(F)**, bars represent mean ± SEM, $n$=14-21 **(C-F)**, $n$=4-16 **(G-J)**.

These ameliorative effects of intranasal IL-10 treatment on cognitive impairments in LH mice led us to test if IL-10 treatment was also effective in genetic, as opposed to stress-dependent, mouse models that display impaired learning and memory, the mouse model of Fragile X syndrome [*FMR1^-/-* mice; (Franklin et al., 2014)] and mice expressing constitutively active glycogen synthase kinase-3 (GSK3) [GSK3 knock-in mice; (McManus et al., 2005)]. IL-10 treatment reversed novel object recognition impairments in both *FMR1^-/-* mice (Fig. 15C-F) and GSK3 knock-in mice (Fig. 15G-J). To investigate possible sex differences within the learned helplessness-dependent impairment, as well as IL-10-dependent rescue, of cognition, we also tested if IL-10 treatment ameliorated NOR deficits in female wild-type mice. Similar to wild-type male mice, IL-10 also reversed novel object recognition impairments in female mice (Fig. 16). Both female NLH and LH mice exhibited novel object recognition impairments, suggesting that learning and memory in females may be more susceptible to the effects of stress than in males while nonetheless proving amenable to the beneficial actions of IL-10 in either case.

**Figure 16. Novel object recognition deficits were observed in both NLH and LH female mice and were rescued in both by IL-10.** Female wild-type mice were subjected or not (NS) to the learned helplessness paradigm and separated into 2 groups: non-learned helpless (NLH) and learned helpless (LH) mice, according to their number of failures out of 30 escapable shock trials and the next day learning and memory was assessed. After the first cognitive assessment, mice were treated with IL-10 (5 μg/mouse) i.n. and retreated 1 h before the learning and memory re-assessment on the next day as shown in **(A)**. The percentage of a mouse's total object exploration time spent exploring the familiar (F) and novel (N) objects **(B-C)** and the total time (s) mice

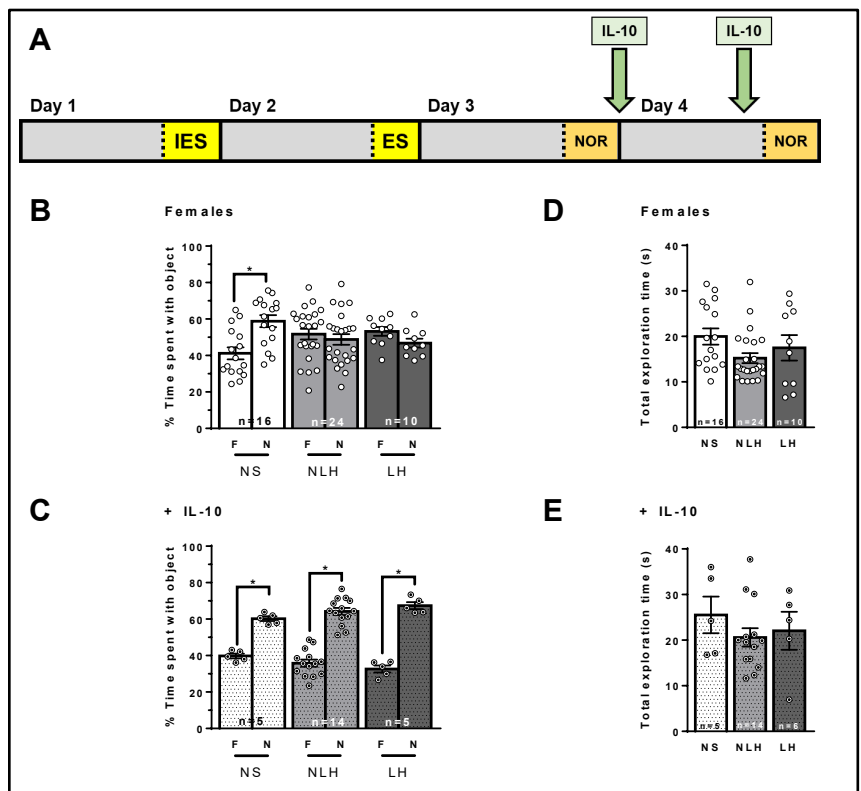

spent engaged in exploration of both the familiar and novel objects combined **(D-E)** in the novel object recognition task were reported. Each dot represents a mouse. Two-way ANOVA, $F_{(2,94)interaction}=7.794$, $F_{(1,94)treatment}=1.019$, $F_{(2,94)condition}=4.685\times10^{-3}$ **(B)**, $F_{(2,42)interaction}=3.306$, $F_{(1,42)treatment}=192.2$, $F_{(2,42)condition}=1.922\times10^{-10}$ **(C)**, Bonferroni post-hoc test, *$p<0.05$, compared to % time spent with familiar object, bars represent mean ± SEM, $n$=5-24 mice/group.

### 4.7 IL-10 was sufficient to rescue the non-learned helplessness-dependent novel object recognition and spatial memory impairments associated with microglial depletion

To determine if the amelioration of cognitive impairments by of the actions of IL-10 requires microglial cells, microglia were eliminated by PLX5622 treatment followed by exposure to the LH paradigm and IL-10 treatment (Fig. 17A). IL-10 administration had no effect on novel object recognition and spatial memory of PLX5622-treated NS mice, or in PLX5622-treated LH mice since the absence of microglia was sufficient to rescue novel object recognition and spatial memory of LH mice (Fig. 17B-D, Fig. 11A). However, in microglia-depleted NLH mice, IL-10 rescued impaired novel object recognition and spatial memory (Fig. 17B-D) as well as hippocampal dendritic spine density (Fig. 9D-F), suggesting that in NLH mice, microglial cells might secrete IL-10 to maintain cognitive performances, consistent with data shown in Fig. 6K. In contrast, in LH mice, the contribution of microglial cells in regulating learning and memory is deleterious, and IL-10 improved learning and memory whether or not microglial cells are present, suggesting that either microglial cells adopt a pro-inflammatory profile in LH mice, and the addition of the anti-inflammatory cytokine IL-10 attenuates the microglial inflammatory response, or that IL-10 acts independently of microglial cells.

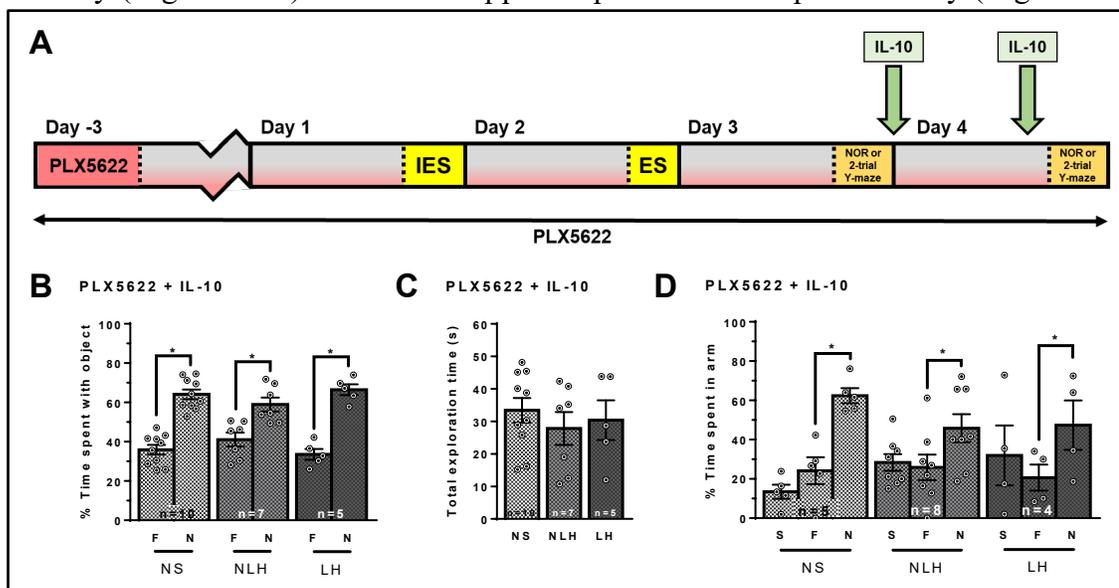

**Figure 17. IL-10 reversed the learning and memory impairments induced by microglia depletion in non-learned helpless mice.** Male mice were fed with the AIN-76A diet supplemented with PLX5622 for 3 days before the induction of the learned helplessness paradigm and were maintained on the diet for the duration of the experiment. Three days after initiating the diet, mice were subjected or not (NS) to the learned helplessness paradigm and separated into 2 groups: non-learned helpless (NLH) and learned helpless (LH) mice, according to their number of failures out of 30 escapable shock trials, and the next day learning and memory was assessed. After the cognitive assessments, mice were treated with IL-10 (5 μg/mouse) i.n. and retreated 1 h before the learning and memory re-assessment on the next day as shown in **(A)**. The percentage of a mouse's total object exploration time spent exploring the familiar (F) and novel (N) objects **(B)** and the total time (s) mice spent engaged in exploration of both the familiar and novel objects combined **(C)** in the novel object recognition test were reported. Each dot represents a mouse. Two-way ANOVA $F_{(2,38)\text{interaction}}=2.918$, $F_{(1,38)\text{treatment}}=109.4$, $F_{(2,38)\text{condition}}=1.415\times10^{-14}$, Bonferroni post-hoc test, $*p<0.05$, compared to % time spent with familiar object, bars represent mean ± SEM, $n=5$-10 mice/group. **(D)** The percentage of a mouse's total maze exploration time spent exploring the start (S), familiar (F), and novel (N) arms in the two-trial Y-maze test was measured in a different cohort of mice than B. Each dot represents a mouse. Two-way ANOVA, $F_{(4,42)\text{interaction}}=1.576$, $F_{(2,42)\text{treatment}}=13.10$, $F_{(2,42)\text{condition}}=1.181\times10^{-14}$, Fisher's least significant difference test, $*p<0.05$, bars represent mean ± SEM, $n=4$-8 mice/group.

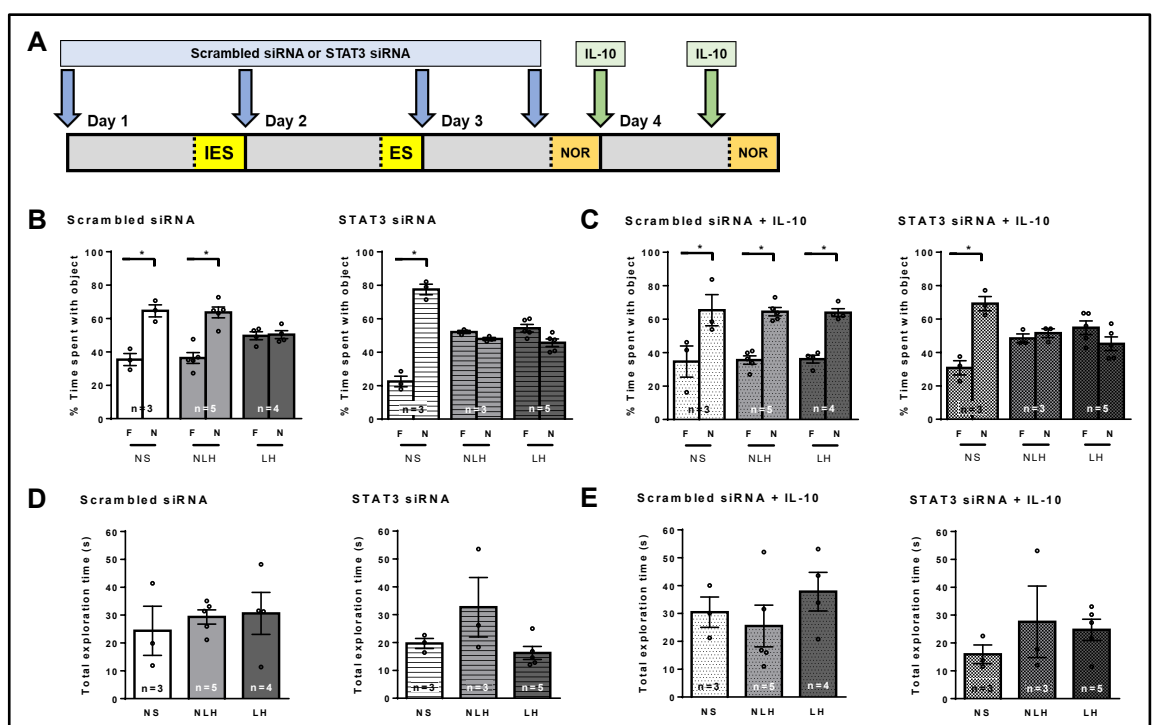

**Figure 18. STAT3 knockdown impaired NOR in NLH mice and blocked IL-10 effects to reverse impairments.** Male mice were treated intranasally with a scrambled siRNA control or STAT3 siRNA (10μg/mouse) 1 day prior to the induction of the LH paradigm and treated every day thereafter. Mice were subjected or not (NS) to the LH paradigm and separated into 2 groups: non-learned helpless (NLH) and learned helpless (LH) mice, according to their number of failures out of 30 escapable shock trials, and the next day NOR was assessed. After the cognitive assessments, mice were treated with IL-10 (5 μg/mouse) i.n. and retreated 1 h before the second NOR re-assessment on the next day as shown in **(A)**. The percentage of a mouse's total object exploration time spent exploring the familiar (F) and novel (N) objects in scrambled siRNA and STAT3 siRNA treated mice and in IL-10-treated, scrambled siRNA- and STAT3 siRNA- treated mice **(B)** as well as the total time (s) mice spent engaged in exploration of both the familiar and novel objects combined **(C)** in the NOR test were reported. Each dot represents a mouse. Two-way ANOVA $F_{(14,68)interaction}$=10.42, $F_{(7,68)treatment}$=24.11, $F_{(2,68)condition}$=3.088×10$^{-13}$, Bonferroni post-hoc test, *$p$<0.05, compared to % time spent with familiar object, bars represent mean ± SEM, $n$=3-5 mice/group.

## 4.8    Depletion of STAT3 abolished the beneficial effect of IL-10

To determine if STAT3 contributes to the IL-10 pro-cognitive effects, since STAT3 mediates IL-10 signaling, we knocked down STAT3 in the brain (Fig. 18A), using STAT3 siRNA delivered intranasally, which we previously showed was sufficient to knock down protein expression in the hippocampus by 75% (Pardo et al., 2017). Seventy-four percent depletion of STAT3 was confirmed by immunohistochemistry (Fig. 19A-H). As expected, knockdown of STAT3 induced novel object recognition impairment in NLH mice (Fig. 18B, C and Fig. 19I), similarly to the microglial depletion by PLX5622. We found that depletion of STAT3 blocked the rescue by IL-10 of the novel object recognition impairment of LH mice, whereas delivery of a scrambled siRNA did not affect IL-10 effects, demonstrating that STAT3 was required to mediate IL-10 effects in LH mice (Fig. 18B, C and Fig. 19I). We also found that the impairment of novel object recognition in NLH mice could not be rescued by IL-10, consistent with the idea that in NLH mice, microglial cells might secrete IL-10 to maintain cognitive performances, and that in the absence of STAT3, IL-10 is unable to mediate its beneficial cognitive effects. However, depletion of STAT3 in NS mice had no effect on novel object recognition, suggesting that IL-10 is critical in the stress response, and that other pathways contribute to learning and memory in non-shocked mice, and might compensate for the absence of STAT3.

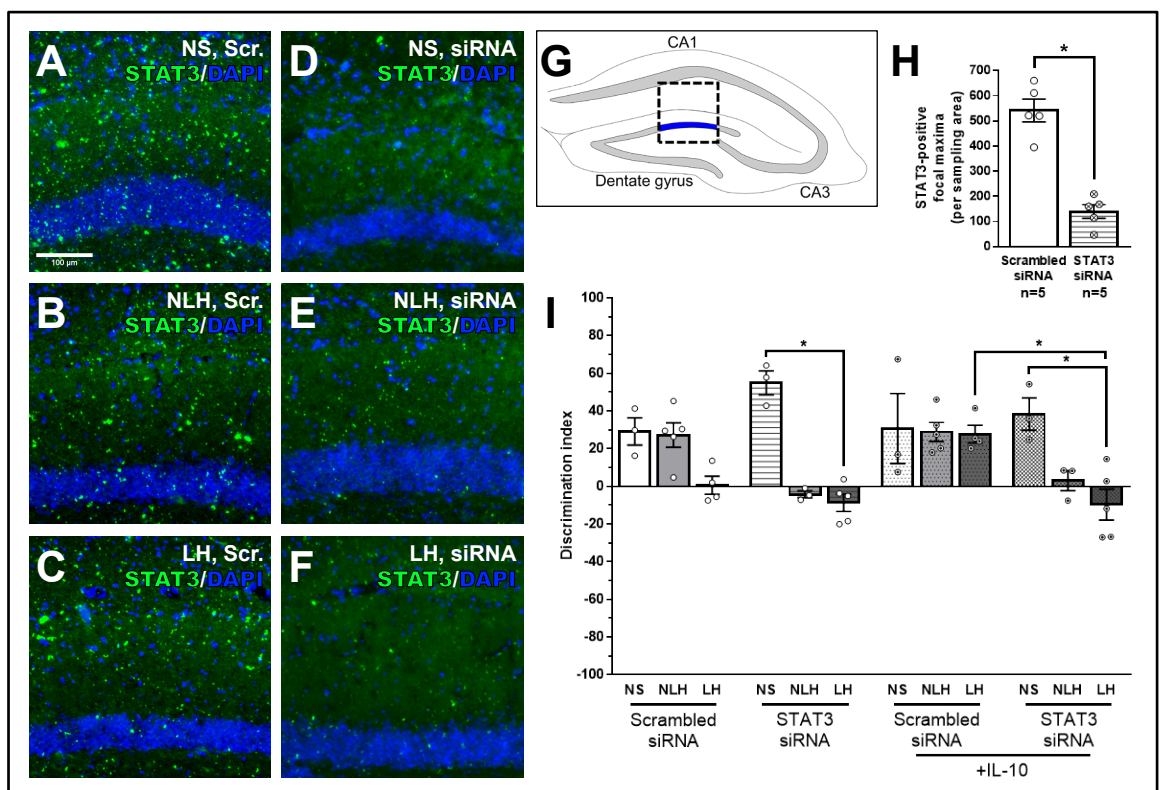

**Figure 19. STAT3 expression was reduced by ~74% in the hippocampus after intranasal treatment with STAT3 siRNA.** Mice were treated intranasally once daily with a scrambled siRNA control or STAT3 siRNA (10 μg/mouse) 1 day prior to the induction of the learned helplessness paradigm and treated every day thereafter. Mice were subjected or not (NS) to the learned helplessness paradigm and separated into 2 groups: non-learned helpless (NLH) and learned helpless (LH) mice, according to their number of failures out of 30 escapable shock trials. One hour after the last treatment, mice were perfused and brains were collected for immunostaining for STAT3 and DAPI. **(A-F)** Representative images of STAT3 immunofluorescence in the hippocampus of scrambled siRNA-treated NS **(A)**, NLH **(B)**, and LH **(C)** mice, and STAT3 siRNA-treated NS **(D)**, NLH **(E)**, and LH **(F)** mice from the region delimited by the square in **(G)**. **(H)** Number of STAT3$^+$ focal maxima per 250,000 μm² sampling area within the hippocampus was quantified. Data from NS, NLH, and LH mice was pooled under control and experimental conditions. Each dot represents a mouse. Mann-Whitney test, U=0, *$p<0.05$, bars represent mean ± SEM, $n$=5/group. **(I)** Discrimination index in the novel object recognition. Each dot represents a mouse. Two-way ANOVA, F(6,34)$_{interaction}$=4.392, F(3,34)$_{treatment}$=3.538, F(2,34)$_{condition}$=22.82, Bonferroni post-hoc test, *$p<0.05$, bars represent mean ± SEM, $n$=3-5 mice/group.

**Table 3. Relevant values from statistical analyses performed on data contained within each figure.**

| Figure 6 Panel | F values | P values | Figure 13 Panel | F values | P values |
|---|---|---|---|---|---|
| F | $F_{(2,17)\text{interaction}} = 9.498$ | $p < 0.05$ | B | $F_{(2,24)\text{interaction}} = 7.334$ | $p < 0.05$ |
|  | $F_{(1,17)\text{treatment}} = 17.09$ | $p < 0.05$ |  | $F_{(1,24)\text{treatment}} = 15.83$ | $p < 0.05$ |
|  | $F_{(2,17)\text{condition}} = 4.181$ | $p < 0.05$ |  | $F_{(2,24)\text{condition}} = 5.380 \times 10^{-14}$ | $p > 0.9999$ |
| G | $F_{(2,17)\text{interaction}} = 13.99$ | $p < 0.05$ | C | $F_{(2,\ 12)} = 0.6711$ | $p = 0.5293$ |
|  | $F_{(1,17)\text{treatment}} = 32.02$ | $p < 0.05$ | D | $F_{(4,36)\text{interaction}} = 14.31$ | $p < 0.05$ |
|  | $F_{(2,17)\text{condition}} = 8.281$ | $p < 0.05$ |  | $F_{(2,36)\text{treatment}} = 86.31$ | $p < 0.05$ |
| K | $F_{(2,\ 21)} = 8.895$ | $p < 0.05$ |  | $F_{(2,36)\text{condition}} = 0.0$ | $p > 0.9999$ |
| **Figure 7 Panel** | **Mann-Whitney U** | **P values** | E | $F_{(2,52)\text{interaction}} = 16.28$ | $p < 0.05$ |
| B | $U = 0$ | $p < 0.05$ |  | $F_{(1,52)\text{treatment}} = 55.27$ | $p < 0.05$ |
| C | $U = 4$ | $p < 0.05$ |  | $F_{(2,52)\text{condition}} = 1.193 \times 10^{-13}$ | $p > 0.9999$ |
| D | $U = 3$ | $p < 0.05$ | F | $F_{(2,\ 26)} = 0.7449$ | $p = 0.4846$ |
| **Figure 8 Panel** | **F values** | **P values** | G | $F_{(4,81)\text{interaction}} = 5.807$ | $p < 0.05$ |
| B | $F_{(2,21)\text{interaction}} = 13.13$ | $p < 0.05$ |  | $F_{(2,81)\text{treatment}} = 9.980$ | $p < 0.05$ |
|  | $F_{(1,21)\text{treatment}} = 47.51$ | $p < 0.05$ |  | $F_{(2,81)\text{condition}} = 0.0$ | $p > 0.9999$ |
|  | $F_{(2,21)\text{condition}} = 6.920$ | $p < 0.05$ | **Figure 14 Panel** | **F values** | **P values** |
| **Figure 9 Panel** | **F values** | **P values** | B | $F_{(2,74)\text{interaction}} = 0.6239$ | $p = 0.5386$ |
| B | $F_{(2,21)\text{interaction}} = 9.738$ | $p < 0.05$ |  | $F_{(1,74)\text{treatment}} = 102.8$ | $p < 0.05$ |
|  | $F_{(1,21)\text{treatment}} = 29.09$ | $p < 0.05$ |  | $F_{(2,74)\text{condition}} = 4.298 \times 10^{-14}$ | $p > 0.9999$ |
|  | $F_{(2,21)\text{condition}} = 5.171$ | $p < 0.05$ | C | $F_{(2,\ 37)} = 2.683$ | $p = 0.0816$ |
| C | $F_{(2,21)\text{interaction}} = 7.890$ | $p < 0.05$ | D | $F_{(4,57)\text{interaction}} = 1.501$ | $p = 0.2141$ |
|  | $F_{(1,21)\text{treatment}} = 46.43$ | $p < 0.05$ |  | $F_{(2,57)\text{treatment}} = 46.50$ | $p < 0.05$ |
|  | $F_{(2,21)\text{condition}} = 3.870$ | $p < 0.05$ |  | $F_{(2,57)\text{condition}} = 4.036 \times 10^{-11}$ | $p > 0.9999$ |
| D | $F_{(2,24)\text{interaction}} = 5.008$ | $p < 0.05$ | **Figure 15 Panel** | **Test statistic** | **P values** |
|  | $F_{(1,24)\text{treatment}} = 15.85$ | $p < 0.05$ | A | Mann-Whitney $U = 17.5$ | $p = 0.2211$ |
|  | $F_{(2,24)\text{condition}} = 8.301$ | $p < 0.05$ | C | $F_{(3,72)} = 11.56$ | $p < 0.05$ |
| E | $F_{(2,24)\text{interaction}} = 14.93$ | $p < 0.05$ | E | $t = 1.452,\ df=37$ | $p = 0.1549$ |
|  | $F_{(1,24)\text{treatment}} = 11.60$ | $p < 0.05$ | D | $F_{(3,60)} = 29.64$ | $p < 0.05$ |
|  | $F_{(2,24)\text{condition}} = 11.36$ | $p < 0.05$ | F | $t = 2.629,\ df=30$ | $p < 0.05$ |
| F | $F_{(2,24)\text{interaction}} = 44.31$ | $p < 0.05$ | G | $F_{(3,46)} = 4.436$ | $p < 0.05$ |
|  | $F_{(1,24)\text{treatment}} = 17.43$ | $p < 0.05$ | I | $t = 0.8193,\ df=23$ | $p = 0.4210$ |
|  | $F_{(2,24)\text{condition}} = 12.09$ | $p < 0.05$ | H | $F_{(3,18)} = 12.42$ | $p < 0.05$ |
| **Figure 10 Panel** | **F values** | **P values** | J | $t = 1.928,\ df=9$ | $p = 0.0859$ |
| A | $F_{(2,90)\text{interaction}} = 15.53$ | $p < 0.05$ | **Figure 16 Panel** | **F values** | **P values** |
|  | $F_{(1,90)\text{treatment}} = 56.16$ | $p < 0.05$ | B | $F_{(2,94)\text{interaction}} = 7.794$ | $p < 0.05$ |
|  | $F_{(2,90)\text{condition}} = 4.238 \times 10^{-14}$ | $p > 0.9999$ |  | $F_{(1,94)\text{treatment}} = 1.019$ | $p = 0.3154$ |
| C | $F_{(2,\ 45)} = 0.1032$ | $p = 0.9021$ |  | $F_{(2,94)\text{condition}} = 4.685 \times 10^{-3}$ | $p = 0.9953$ |
| D | $F_{(4,84)\text{interaction}} = 5.373$ | $p < 0.05$ | D | $F_{(2,\ 47)} = 2.393$ | $p = 0.1024$ |
|  | $F_{(2,84)\text{treatment}} = 11.68$ | $p < 0.05$ | C | $F_{(2,42)\text{interaction}} = 3.306$ | $p < 0.05$ |
|  | $F_{(2,84)\text{condition}} = 7.654 \times 10^{-14}$ | $p > 0.9999$ |  | $F_{(1,42)\text{treatment}} = 192.2$ | $p < 0.05$ |
| **Figure 11 Panel** | **F values** | **P values** |  | $F_{(2,42)\text{condition}} = 1.922 \times 10^{-10}$ | $p > 0.9999$ |
| A | $F_{(2,82)\text{interaction}} = 4.627$ |  | E | $F_{(2,\ 21)} = 0.6662$ | $p = 0.5242$ |
|  | $F_{(1,82)\text{treatment}} = 2.914$ | $p = 0.0916$ | **Figure 17 Panel** | **F values** | **P values** |
|  | $F_{(2,82)\text{condition}} = 2.991$ | $p = 0.0557$ | B | $F_{(2,38)\text{interaction}} = 2.918$ | $p = 0.0663$ |
| A | $F_{(4,57)\text{interaction}} = 4.832$ |  |  | $F_{(1,38)\text{treatment}} = 109.4$ | $p < 0.05$ |
|  | $F_{(2,57)\text{treatment}} = 1.984$ | $p = 0.1468$ |  | $F_{(2,38)\text{condition}} = 1.415 \times 10^{-14}$ | $p > 0.9999$ |
|  | $F_{(2,57)\text{condition}} = 3.043$ | $p = 0.0555$ | C | $F_{(2,19)} = 0.3981$ | $p = 0.6771$ |
| C | $F_{(4,24)\text{interaction}} = 3.685$ | $p = 0.0178$ | D | $F_{(4,42)\text{interaction}} = 1.576$ | $p = 0.1985$ |
|  | $F_{(2,24)\text{treatment}} = 68.48$ | $p < 0.05$ |  | $F_{(2,42)\text{treatment}} = 13.10$ | $p < 0.05$ |
|  | $F_{(2,24)\text{condition}} = 8.465 \times 10^{-11}$ | $p > 0.9999$ |  | $F_{(2,42)\text{condition}} = 1.181 \times 10^{-14}$ | $p > 0.9999$ |
| D | $F_{(4,18)\text{interaction}} = 2.102$ | $p = 0.1228$ | **Figure 18 Panel** | **F values** | **P values** |
|  | $F_{(2,18)\text{treatment}} = 70.25$ | $p < 0.05$ | B | $F_{(14,68)\text{interaction}} = 10.42$ | $p < 0.05$ |
|  | $F_{(2,18)\text{condition}} = 1.901 \times 10^{-13}$ | $p > 0.9999$ |  | $F_{(7,68)\text{treatment}} = 24.11$ | $p < 0.05$ |
| F | $F_{(4,27)\text{interaction}} = 2.219$ | $p = 0.0935$ |  | $F_{(2,68)\text{condition}} = 3.088 \times 10^{-13}$ | $p > 0.9999$ |
|  | $F_{(2,27)\text{treatment}} = 107.1$ | $p < 0.05$ | C | $F_{(6,34)\text{interaction}} = 1.035$ | $p = 0.4201$ |
|  | $F_{(2,27)\text{condition}} = 7.569 \times 10^{-11}$ | $p > 0.9999$ |  | $F_{(3,34)\text{treatment}} = 1.221$ | $p = 0.3171$ |
| G | $F_{(4,21)\text{interaction}} = 1.188$ | $p = 0.5857$ |  | $F_{(2,34)\text{condition}} = 0.8711$ | $p = 0.4276$ |
|  | $F_{(2,21)\text{treatment}} = 58.12$ | $p < 0.05$ | **Figure 19 Panel** | **Test statistic** | **P values** |
|  | $F_{(2,21)\text{condition}} = 8.575 \times 10^{-14}$ | $p > 0.9999$ | H | Mann-Whitney $U = 0$ | $p < 0.05$ |
| **Figure 12 Panel** | **t test** | **P values** | I | $F_{(6,34)\text{interaction}} = 4.392$ | $p < 0.05$ |
| A | $t = 13.30,\ df=4$ | $p < 0.05$ |  | $F_{(3,34)\text{treatment}} = 3.538$ | $p < 0.05$ |
| B | Mann-Whitney $U = 88$ | $p = 0.1403$ |  | $F_{(2,34)\text{condition}} = 22.82$ | $p < 0.05$ |

## Chapter 5 – Discussion

### 5.1    Summary of findings

The mechanisms of cognitive impairments induced by stress and those associated with depression remain poorly understood. We found that the induction of learned helplessness in mice was associated with impaired novel object recognition and spatial memory, decreased hippocampal dendritic spine density, increased microglial activation and slightly reduced IL-10-producing microglia. Microglial cells were found to be involved in the novel object recognition and spatial memory impairments because depleting microglia prevented the development of cognitive impairments in learned helpless mice. We further identified the IL-10/STAT3 pathway as a potential mechanism promoting novel object recognition and spatial memory after stress.

### 5.2    Evidence for the involvement of microglia and IL-10 in learning and memory pathology after stress

Consistent with previous reports, microglial cells were activated in response to stress, and this correlated with reduced dendritic spine density (Wohleb et al., 2018). Microglial cells are able to phagocytose synapses, and therefore shape neuronal circuits (Dantzer et al., 2008; Schafer et al., 2012; Wohleb, 2016). Indeed, resting microglial cells extend and retract their processes to actively survey their environment, including making contacts with neurons (Wake et al., 2009). This is thought to have beneficial consequences, such as eliminating ischemic synapses and refining the wiring of neurons (Brown & Neher, 2014). We found that microglial depletion enhances learning and memory after stress. However, it is difficult to detangle if the improvement of learning and memory after PLX5622 could also result from an amelioration of depressive-like behavior, as PLX5622 induces resilience to learned helplessness induction. Nevertheless, it is noticeable that PLX5622 alters learning and memory in resilient mice while ameliorating learning and memory in susceptible mice, uncovering the complex role of microglial cells in learning and memory after stress. In pathological conditions microglia often adopt an amoeboid phenotype and migrate to sites of injury where they release factors that have been reported to result in either neuroprotection or neurotoxicity (Biber et al., 2007; Vilhardt, 2005). Based on our findings, one of these factors appears to be IL-10. We found that after stress, microglial cells from LH mice express slightly less IL-10 than microglia from NLH mice. This corroborates findings that homeostatic and disease-related microglia may be two distinct populations (Hammond et al., 2019). Consistent with this, IL-10 levels have been shown to be diminished in depression (Dhabhar et al., 2009), and a defective anti-inflammatory IL-10 pathway has been associated with resistance to antidepressant treatments (Syed et al., 2018). This suggests that some microglial cells might have anti-inflammatory properties in response to stress, participating in the promotion of learning and memory in stressed mice, whereas in stressed LH mice, activated microglial cells acquire an inflammatory phenotype, leading to a reduction of the beneficial action of IL-10 on learning and memory. It is also possible that a new population of microglial cells is induced in LH mice that directly impair learning and memory. To reinforce the idea that microglia are critical in maintaining learning and memory after stress, we found that mice that received foot-shocks but were resilient to learned helplessness induction exhibited cognitive impairments and dampening of dendritic spine density after microglial depletion (Fig. 9D-F). This was rescued by the administration of IL-10, suggesting that 1) microglia are required for learning and memory in stress conditions, even if the mice are resilient, and 2) the microglial cells in NLH mice produce the required IL-10 to maintain learning and memory. Whether the loss of IL-10 expression occurs in selective populations of microglia or results from the induction of a novel population of disease-related microglia in which the production of IL-10 is inhibited, or both, remains to be determined. It is also possible that IL-10 is produced by other cells in the brain than microglial cells but that IL-10 production by other cells requires signaling from the microglia. IL-10 might have direct effects on microglia to prevent them from phagocytosing synapses (Saijo & Glass, 2011) and/or IL-10 might also act directly on neurons to increase synapse formation (Lobo-Silva et al., 2016; Zhou et al., 2009b). Furthermore, the mechanisms whereby IL-10 promotes dendritic spine density remain to be elucidated.

### 5.3    STAT3 as a mediator for IL-10 effects

It was reported that STAT3 is involved in synapse formation (Su et al., 2019), so IL-10 might promote dendritic spine density by activating STAT3. We showed that STAT3 mediates the behavioral effect of IL-10, as depletion of STAT3 was sufficient to block IL-10 rescue of novel object recognition and spatial memory impairments in LH mice. This adds to the literature showing that STAT3 is important to mediate NMDAR-LTD to modulate synaptic plasticity (Nicolas et al., 2012). Therefore, whether this is the mechanism whereby IL-10 promotes learning and memory in learned helpless mice remains to be further studied. However, it is important to note that depletion of STAT3 in NLH mice but not NS mice also impairs learning and memory, confirming that STAT3 promotes learning and memory in response to stress and favors resilience. Other pathways inducing STAT3 activation might also contribute to learning and memory after stress, but this will need further investigations.

### 5.4    IL-10's anti-inflammatory action

The reduction in IL-10 levels or blockade in IL-10 signaling might in part also explain the puzzling findings that pro-inflammatory cytokines are required at physiological levels to maintain synaptic plasticity and proper neuronal network functioning (Benarroch, 2013; Brynskikh et al., 2008), yet excess levels of pro-inflammatory cytokines are associated with impairments of learning and memory (Dantzer et al., 2008; McAfoose & Baune, 2009). It is indeed possible that in the presence of high levels of pro-inflammatory cytokines, IL-10 production or its actions are suppressed, abolishing IL-10 associated benefits for learning and memory, instead of direct alteration of neuronal circuits from pro-inflammatory cytokines.

Dendritic spine density perturbations in transgenic $FMR1^{-/-}$ (FX) mice compared to wild-type appear to be variable between studies, especially in the hippocampus. However, evidence suggests that impaired stability and increased turnover leads to a decrease in *functional* hippocampal dendritic spines in FX mice, suggesting FX and LH mice may share dendritic spine dysfunction as a common pathological source for their cognitive abnormalities (He & Portera-Cailliau, 2013). The pathophysiology of the FX phenotype was previously shown to include GSK3 hyperactivity, which would be expected to constrain anti-inflammatory activity. It is noticeable that IL-10 exhibited pro-cognitive effects in FX and GSK3 knock-in mice without stress, suggesting that similar pathways contribute to learning and memory impairments in these mice as in LH mice, and that IL-10 is able to reverse these effects, pointing toward GSK3 action in impairing learning and memory in FX mice, as previously reported (Franklin et al., 2014; King &

Jope, 2013). Further experiments will be needed to determine the mechanism whereby IL-10 improves learning and memory in FX and GSK3 knock-in mice.

Altogether my findings show that IL-10 provides pro-cognitive actions in LH mice or mice with cognitive impairments possibly associated with microglial dysfunction affecting synapse balance. Further, learned helplessness may be associated with an as-yet uncharacterized pathological subtype of microglia, deficient in IL-10 production, which could impart vulnerability to stress. Dysregulated microglial activation in response to stress has the potential to damage or disrupt neural circuit integrity in brain regions critical for learning and memory, like the hippocampus, and accordingly impair cognitive functions that network integrity underlies. On the other hand, stress resilience in NLH mice may be associated with neuroprotective microglia which, upon stress-induced activation, produce IL-10 in amounts sufficient to preserve functional neural network integration, protecting learning and memory functions. Depleting NLH mice of their microglia with PLX5622 would consequently remove the pro-cognitive benefit they provide, imperiling neuronal connectivity and any contingent learning and memory functions (Fig. 20). These potential pathways to learning and memory deficits, mediated by microglia which critically differ in the IL-10-associated protection they provide, offer novel promising avenues of investigation for therapies aiming to correct learning and memory deficits.

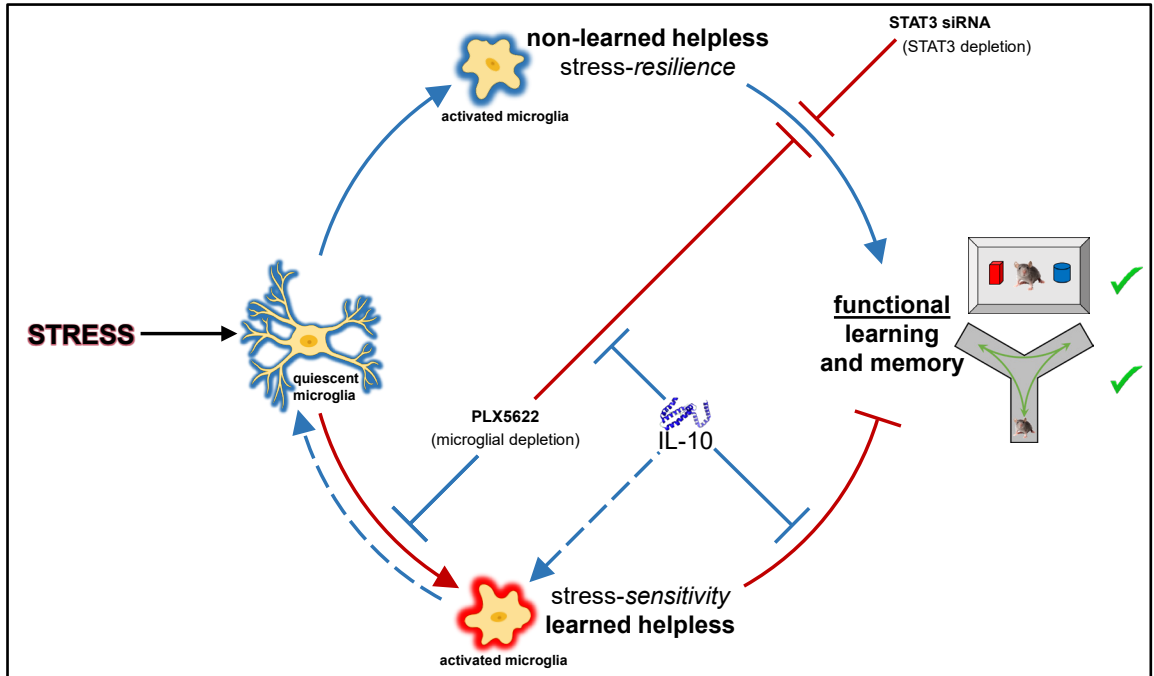

**Figure 20. Schematic representation of the role of microglial cells and IL-10 in learning and memory impairments associated with sensitivity to stress as well as microglial depletion.**

# Chapter 6 – Implications for the Field

## 6.1 Development of a novel model to understand the effect of stress-induced depressive-like behaviors on learning and memory

In this study, utilizing the learned helplessness paradigm, we developed a model of stress-induced cognitive impairment associated with depressive-like behavior in mice. Because the stress (in the form of foot-shocks) applied to animals, as well as the test to measure the development of a depressive phenotype, is condensed into a two-day span, our model is a relatively rapid and streamlined method to pair the measurement of a memory modality of interest with the measurement of an animal's emotion-like response to the stress. This model creates a closed system that includes animal subjects in a pre-disease state, disease state with no therapeutic intervention, and a post-therapy state. Because of the acute nature of the pathogenic stimulus, clear distinctions can be drawn from measurements made during each of these three physiological states, allowing for a more comprehensive characterization of disease and therapeutic mechanisms. No prior training of the mice or manipulations to their physiology are necessary, avoiding a significant confound of other models that makes such comparisons more difficult. By conducting a baseline measurement of learning and memory (after escapable shocks (ES) but before treatment administration), with a second round of learning and memory tests after treatment using the same cohort, physiological correlations can be established that link a given therapy with changes to learning and memory performance.

The field of depression research has long been in need of models of cognitive impairment associated with depression, as evidenced by the great difficulty in addressing commonly persistent and refractory symptoms within the cognitive domain, often despite effective clinical management of mood symptoms. My model demonstrates that the induction of learned helplessness is sufficient to induce cognitive impairment, further adding to the validity of the learned helplessness paradigm to recapitulate symptoms of depression experienced by humans. Thus, this model provides new avenues for investigation into the effect of stress on learning and memory, in particular probing the potential role of neuroinflammation in mediating stress-induced cognitive impairments that are reversed by the action of IL-10. Despite excellent specificity, and face and predictive validity, this model of LH-dependent impaired cognition has limitations. Aside from limits to accessibility due to the need for specialized equipment, the major limitation of the learned helplessness paradigm is that depressive-like symptoms that are generated after inescapable foot-shocks (IES) eventually resolve spontaneously, limiting the time window within which depressive-like behavior can be studied to roughly no more than two weeks following IES.

## 6.2 Intranasal delivery of anti-inflammatory cytokines demonstrates potential to reduce neuroinflammation

We demonstrated the therapeutic potential of anti-inflammatory cytokine treatment to improve learning and memory impairments associated with stress-induced depressive-like behavior. We also confirmed the hypothesis that a depressive phenotype was associated with evidence in the hippocampus of an exaggerated neuroinflammatory response to stress, and that IL-10, which has well-studied anti-inflammatory effects, was sufficient to reverse microglial activation in the hippocampus as well as improve learning and memory in impaired, LH animals.

Additionally, we established the viability of intranasal administration of anti-inflammatory cytokines directly to the brain to elicit effects within the brain, highlighting this drug delivery method as a feasible clinical tool for targeting the CNS while bypassing the circulatory system, the BBB, and the risk of protein degradation associated with other delivery methods. Many small proteins of similar or greater molecular weight to IL-10 (MW=18.7 kDa), including radio-labeled human: insulin-like growth factor-1 ([125I]-IGF-1) (MW=7.65 kDa), erythropoietin (MW=21 kDa), transforming growth factor (TGF)-β (MW=25 kDa), nerve growth factor (NGF) (MW=26.5 kDa), vascular endothelial growth factor (VEGF) (MW=38.2 kDa), and glucagon like peptide-1 (GLP-1) (MW=59.7 kDa), have been shown to reach the hippocampus following intranasal administration and to yield significantly higher CNS concentrations than comparable intravenous or intraperitoneal dosing in rodents and monkeys (Chauhan & Chauhan, 2015; Lochhead & Thorne, 2012). Transport of small peptides, such as IL-10, across the olfactory epithelium can occur by a paracellular transport mechanism in which the peptide leaks through the intercellular spaces or a transcellular mechanism whereby the peptide is transported through cells by specific channels and pores, transcytosis, and by transcellular diffusion (Lochhead & Thorne, 2012; Meredith et al., 2015). Peptides then travel along ethmoidal nerves across the cribriform plate to arrive at the olfactory bulb in under 30 minutes following administration via rapid bulk flow, and in 1 to 3 hours via fast and slow axonal transport or diffusion (Mittal et al., 2014; Renner et al., 2012). A peptide's spread throughout the rest of the brain is mediated via more rapid bulk flow, i.e. extracellular convection within the perivascular spaces of cerebral blood vessels, as well as slower intracellular axonal transport along with extracellular diffusion along synapsing neurons, reaching a wide distribution in a matter of hours.(Chauhan & Chauhan, 2015; Lochhead & Thorne, 2012). It would be reasonable to expect that, insofar as IL-10 and other anti-inflammatory cytokines have similar molecular properties to those of the small peptides that have been measured, intranasal delivery of IL-10 and other anti-inflammatory cytokines would result in a similar distribution throughout the brain, as these cellular transport mechanisms have been thoroughly characterized and are well-understood. Thus, additional anti-inflammatory candidates can be tested to determine their action in comparison to IL-10 administration.

Finally, IL-10 demonstrated therapeutic effects that were apparent 24 hours after the first treatment, an effect onset that is significantly more rapid than that achieved by any currently used antidepressants other than ketamine in humans. Thus, our model represents an important proof-of-concept for intranasal delivery of therapies targeting neuroinflammation as a means of quickly modulating brain activity and alleviating symptoms associated with MDD, like impaired cognition.

### 6.3    Confirmation that microglial cells have various phenotypes

We presented evidence for the existence of IL-10-producing microglia that merit further investigation as a potential determinant factor in the pathogenesis of depressive-like behavior, differentiating stress-sensitive mice that become learned helpless from stress-resilient mice that do not. It is well-known that microglia play a critical role in the maturation of neural circuits during brain development through synaptic pruning whereby microglia extend processes and phagocytose dendritic spines, the location of synaptic connections between neurons (Wake et al., 2013). Microglia engage in this type of behavior as a means to refine neural circuitry such that infrequently active connections that would otherwise make communication between neurons inefficient, are eliminated by engulfment, mediated through the recognition of complement component C3 on neurons by CD11b on the cell surface of microglia (Nakanishi, 2014). Additionally, recognition of the neuronal ligand CX3CL1 (fractalkine) by the microglial CX3CR1 receptor targets synapses for phagocytosis during early postnatal synaptic pruning in the hippocampus (Chung et al., 2015). It is not known if this same mechanism guides microglial processes into close association with neuronal elements in the context of stress and neuroinflammation (Wohleb, 2016). Nonetheless, what is no longer an open question is whether microglia activated by pro-inflammatory cytokines engage in removal of synapses post-developmentally in the adult brain (Miyamoto et al., 2013; Wohleb et al., 2018). This type of activity could potentially perturb mood- and cognition-relevant neural circuitry in the striatum, PFC, and limbic structures, like the hippocampus, if not properly regulated and constrained. Thus, cognitive impairments in depression could be associated with an insufficiency in anti-inflammatory signaling following stress, with disease-related microglia deficient in IL-10 production as a possible pathogenic source.

In support of the idea that differential responses to stress characterize distinct microglial phenotypes, evidence gained from single-cell RNA sequencing approaches suggests the existence of various clusters of microglial subpopulations with unique transcriptional signatures both during homeostatic and disease states. In particular, a subset of microglia that selectively expresses the chemokine *Ccl4* was identified, which were typically present in small numbers but expanded in the context of injury. These microglia were also found to express *Il-1b* and *Tnf*, distinguishing them as a specialized collection of cells uniquely primed to produce an inflammatory response (Hammond et al., 2019). It is quite possible that a subpopulation of microglia with these pro-inflammatory characteristics become more active in response to stress and contribute to the initiation of neuroinflammation. Furthermore, the pro-inflammatory cytokines produced by this subpopulation are detrimental to healthy neuronal function and attract infiltrating immune cells (Gadani et al., 2015). It could be that within this usually small microglial subpopulation, differences in degree of activation have the potential either to produce healthy homeostatic or excessive pathological inflammatory responses consistent with what was observed following learned helplessness. Future RNA sequencing approaches to identify transcriptional variations among microglia in NS, NLH, and LH mice should help illuminate if this is the case.

Microglial depletion with PLX5622 rescued learning and memory in LH mice yet led to a learning and memory deficit in NLH mice, and this impairment was reversed following treatment with IL-10. This suggests that some number of microglia in LH mice display pathogenic characteristics while microglia in NLH mice, on the other hand, may generally be protective. Furthermore, a deficit in available IL-10 may be a critical component in the development of stress-associated symptomatology driven by inflammation such as reduced hippocampal dendritic spine density and impaired learning and memory. Because the pro-cognitive effects of IL-10 were apparent under conditions in which microglia were present as well as when microglia were depleted, IL-10 may serve a more prolific and integrated role in the actions of CNS cell-types other than microglia, with the capacity to correct neural dysfunction and restore learning and memory through multiple available and compensatory pathways. An in-depth understanding of the various microglia populations will be necessary before conclusions can be drawn about the role of microglia in the disease processes at work.

### 6.4 Discovery of a stress-dependent regulation of learning and memory distinct from homeostatic learning and memory

Evidence presented suggests that IL-10 plays a critical role in the regulation of learning and memory after stress. The differences in learning and memory, dendritic spine density, and microglial activation between NLH and LH mice imply that the capacity of CNS cells to produce IL-10, thus constraining the degree of neuroinflammation associated with the stress response, is an important protective factor for proper neuronal function within the hippocampus after stress. IL-10 holds the potential to either preserve healthy learning and memory mechanisms, if produced in sufficiently high amounts, or expose animals to the damaging effects of neuroinflammation, if produced in amounts that are insufficient to properly regulate the brain's inflammatory response, leading to disruption in cognitive function. Further, this neuroprotective and pro-cognitive role for IL-10 would presumably differ from its role within learning and memory processes under healthy physiological conditions, becoming critically relevant after stress.

We also found that changes in hippocampal dendritic spine density might be linked to learning and memory ability rather than learned helplessness, as illustrated by the fact that microglia-depleted LH mice had intact dendritic spine density, while exhibiting learned helplessness. This suggests that removal of microglia in LH mice is beneficial to neurons, implying they contribute to the reduction of dendritic spine*s* after stress. In contrast, microglia in NLH mice may be neuroprotective, contributing to dendritic spine *preservation* after stress which is reflected by hippocampal dendritic spine density.

**Chapter 7 – Future Directions**

**7.1 Identification of CNS cells that produce and respond to IL-10**

In the CNS, it is known that under physiological conditions, both microglia and astrocytes produce IL-10 (Park et al., 2006). However, following stress, the main cellular source of IL-10 in the context of learned helplessness, though suspected of being microglia, has not been identified. Furthermore, IL-10 receptors are known to be located on microglia, astrocytes, and neurons (Lobo-Silva et al., 2016) and we provide evidence that IL-10 is capable of activating pathways with therapeutic benefit, even under conditions of microglial depletion, suggesting the participation of non-microglial neural cells in IL-10-dependent, pro-cognitive processes. Thus, determining the primary cellular producers and targets of IL-10 under stressed conditions will provide a greater understanding of the extent to which IL-10 serves crucial neuroprotective roles after the activation of neuroinflammation. It will also allow for a more complete characterization of cell-types involved in the pathogenesis of depressive-like behavior, thus providing insight into potential disease targets for future therapies. This could be achieved using flow cytometry to quantify the number of IL-10 producing cells within various cell-types, or FISH staining associated with immunofluorescence.

Dysfunctional astrocytes may fail in critical neuroprotective roles, and thus promote the development of depression once the neuronal damage that results becomes significant. For instance, astrocytes support proper neuronal function by taking up synaptic glutamate, which would quickly become excitotoxic leading to neuron cell death if astroglial sequestration of glutamate becomes compromised. Additionally, neurons rely on astrocytes for the production of growth factors, which are essential for branching of neuronal extensions as well as for the repair of cellular damage when it occurs (Sofroniew & Vinters, 2010). Thus, failure in this role would be highly detrimental to neuronal health. Consistent with this, astrocyte density (Miguel-Hidalgo et al., 2000) and synapse-related genes (Kang et al., 2012) are both decreased in the dlPFC of MDD patients. In addition, knockout of the astroglial glutamate transporter GLT-1, which is important for the reuptake of extracellular glutamate, in the lateral habenula of mice, is sufficient to induce depressive-like behaviors (Cui et al., 2014). Altogether, these findings suggest that dysfunction in astrocyte activity can promote depression by impairing neuronal functions and may therefore represent a valuable target for therapy.

Investigation into a possible relationship between astrocytes and microglia regarding IL-10 signaling showed that the main target of IL-10 signaling in the brain may be astrocytes, which in turn exert regulatory control over microglial activity. Astrocytes cultured from brain homogenate expressed higher amounts of IL-10R1 than microglia and, accordingly, demonstrated a higher sensitivity to treatment with IL-10. Additionally, when co-cultured, IL-10 treatment was shown to induce astrocyte production of TGF-β, which in turn induced a shift in the microglia toward an anti-inflammatory profile, reducing microglial expression of IL-1β and increasing expression of anti-inflammatory mediators like fractalkine receptor CX3CR1 and IL-4 receptor α (IL-4Rα) (Norden et al., 2014). This may be yet another way that astroglial atrophy is depressogenic in that, according to this model, loss of astrocytes would result in lower levels of TGF-β produced in response to microglial IL-10, thus halting microglia in an activated, pro-inflammatory state. This interestingly also provides a theoretical context into which our proposed IL-10-deficient microglial disease model fits quite nicely. If the activation state of microglia is regulated by astroglial TGF-β produced in response to IL-10, and microglia in the brains of LH animals secrete insufficient amounts of IL-10 compared to NLH microglia, LH microglia would have less of a regulatory check on their activity, possibly allowing heightened or extended microglial activation that disrupts neuronal functioning.

Finally, it could turn out that, contrary to expectations, the primary source of IL-10 is determined to be peripheral, in which case the pathways activated within IL-10-responsive CNS cell-types would need to be identified.

## 7.2 Molecular mechanism(s) by which IL-10 promotes dendritic spine density and learning and memory

The possibility that IL-10 promotes an increase in dendritic spine density through an effect on microglia, that prevents them from phagocytosing synapses, is supported by evidence we present in this study. The strong correlation between the decrease in markers of microglial activation and the increase in dendritic spine density in LH mice following IL-10 treatment is compelling, but further studies will be necessary to determine the degree to which any direct effect on microglia preserves dendritic spines, as we demonstrated in PLX5622-treated animals that IL-10 is capable of promoting dendritic spine density independently of microglia.. IL-10 treatment is, at minimum, associated with a shift toward microglial quiescence. Depending on the rate of synaptogenesis compared to dendritic spine elimination, at any given time, there exists either a net positive, net negative, or steady state in terms of the overall number of synapses present in the hippocampus. We observed a decrease in dendritic spine density following learned helplessness, compared to non-stressed controls, and propose a theoretical framework in which stress increases neuroinflammation which leads to increased microglial activation, setting the stage in a subset of vulnerable animals for disease-related microglia to engage in aberrant synaptic pruning, likely to a degree that exceeds what is healthy for the animal. If microglial damage to neuronal elements and synapse removal occurs to a high enough degree, this may shift the overall synapse formation-elimination balance in the direction of overall synapse loss. We have shown that IL-10 treatment promotes a shift back to quiescence in activated microglia which may dampen synapse removal to such a degree that synaptogenesis dominates, resulting in a net increase in the number of functional synapses, which in theory could return dendritic spine density to a value similar or exceeding that seen in non-stressed controls.

We also demonstrated that the IL-10-dependent increases in synaptic density can occur independent of microglia. This widens the number of potential mechanisms by which IL-10 signaling could elicit synaptogenesis. If the impairments to learning and memory that have been observed are the result of changes to neuronal connectivity, and IL-10 signaling sufficiently restores functional connectivity to improve learning and memory, since the pro-cognitive benefits associated with IL-10 treatment are still apparent after microglial depletion, it is possible that IL-10 modulates astrocyte activity to influence new synapse formation, as astrocytes express IL-10 receptors. Abnormally elevated levels of pro-inflammatory cytokines have been shown to exert inhibitory pressure on LTP, a process critical for the maintenance and expansion of functional synapses. Thus, IL-10's downstream effects, which may be brought about via multiple pathways including various cell types, should be expected to release the inhibitory pressure placed on LTP by pro-inflammatory signaling, allowing for synapses to strengthen, preventing synapse atrophy, and resulting in an increase in synaptic density.

Additionally, IL-10 may act directly on neurons to precipitate synaptogenesis. IL-10 applied to cultures containing hippocampal neurons and developing microglia without direct contact led to an increase in the numbers of dendritic spines as well as the formation of new synapses (Lim et al., 2013). Furthermore, IL-10 was shown to be capable of acting directly on neurons in culture to promote neurite outgrowth and synapse formation by upregulating expression of Netrin-1 through the JAK1/STAT3 pathway (Chen et al., 2016; Ihara et al., 1997). Moreover, IL-1β exerts an inhibitory effect on glutamate release and LTP and IL-10 reversed this effect by inhibiting IL-1β production and/or release as well as possibly promoting shedding of IL-1 type 1 receptors on cells within the hippocampus (Kelly et al., 2001).

In view of these findings, future investigations will be necessary to determine if any of the aforementioned mechanisms underlie the IL-10-dependent increase in hippocampal dendritic spine density that restores healthy learning and memory in the context of learned helplessness. Conditional cell-type-specific genetic knockouts for proteins suspected of participating in the downstream actions of IL-10 could be utilized to identify the major biomolecular players critically involved in the observed elevations in dendritic spine density.

### 7.3    Characterization of the various glial phenotypes after stress

The observations of reduced dendritic spine density in the hippocampus of depressed humans, as well as LH mice, have yet to be explained in terms of underlying mechanisms at play (see Fig. 1). It is possible that the inhibitory pressure placed on LTP by high levels of inflammation results in an overabundance of weak synapses that either atrophy or are pruned as part of efficiency-saving processes in place to optimize neural networks under otherwise healthy conditions.

Postmortem analysis of brain tissue taken from depressed patients who committed suicide has revealed significant microglial activation in the dlPFC, ACC, mediodorsal thalamus, and hippocampus (Steiner et al., 2008; Steiner et al., 2011). Markers of microglial activation *in vivo*, analyzed using PET of patients during a major depressive episode, were shown to be elevated in the prefrontal cortex, ACC, and insula (Setiawan et al., 2015). Again *in vivo*, correlational fMRI analyses in patients indicate that MDD is associated with alterations in functional connectivity of multiple large-scale networks in the brain (Mulders et al., 2015). Finally, using microarray gene profiling to measure expression of synapse-related genes and electron microscopic stereology to measure number of synapses in depressed suicides, MDD has been linked to a loss of synapses in the dlPFC (Kang et al., 2012). Because microglia have been shown to become activated upon elevated production of cytokines in the periphery and/or CNS and microglial activation in the context of MDD has been shown to occur in many of the same brain regions as altered neural circuitry and loss of synapses in humans, this implicates activated microglia in the direct alteration of neural circuitry associated with MDD (Delpech et al., 2015; Kang et al., 2012; Mulders et al., 2015; Setiawan et al., 2015; Steiner et al., 2008; Steiner et al., 2011).

Evidence gained in this study may suggest a dichotomous role for microglia as either potential neuroprotector or overzealous enforcer of network efficiency depending on how the microglia generally respond to stress and increased neuroinflammation. The impairments displayed by LH mice seem to be associated with IL-10-deficient microglia, which could plausibly be hyperactively removing dendritic spines while in an overactivated, stress-induced state. One way to probe this issue would be to either genetically or pharmacologically disrupt receptor-ligand interactions by which microglia locate, engulf, and remove dendritic spines to measure the degree to which aberrant synaptic pruning plays a part in the observed LH-dependent synaptic losses.

Another psychiatric disorder which is thought, at least in part, to be brought about by aberrant synaptic pruning at the hands of microglia is schizophrenia. Mutations significantly associated with schizophrenia were identified in the genes encoding nitric oxide synthase 1 and thrombospondin 4, which have been linked to microglial activation and phagocytosis in the cortex and thalamus (Neniskyte & Gross, 2017). Accordingly, these two proteins represent two additional targets of interest to test if inhibiting their action reduces the severity of dendritic spine loss following learned helplessness while improving learning and memory. Additionally, microglia may play a role in the pathogenesis of Rett syndrome, a neurodevelopmental disorder characterized by cognitive impairment and loss of various motor functions, through actions that are neurotoxic, possibly originating from abnormalities in microglial glutamate synthesis or release (Maezawa & Jin, 2010).

CNS cell types other than microglia, such as astrocytes, have been implicated in potentially serving pathogenic roles regarding synaptic integrity. Astrocytes are already suspected of playing a role in depression; MDD is associated with loss and hypotrophy of astrocytes (Davis et al., 2002; Miguel-Hidalgo et al., 2010). This mainly occurs in the frontolimbic systems that are relevant for major depression (Altshuler et al., 2010; Cobb et al., 2016; Cotter et al., 2002; Gittins & Harrison, 2011; Medina et al., 2016; Miguel-Hidalgo et al., 2000; Nagy et al., 2015; Ongür et al., 1998; Rajkowska et al., 1999;

Rajkowska & Stockmeier, 2013; Rial et al., 2016; Rubinow et al., 2016; Torres-Platas et al., 2014). Interestingly, there is no astrogliosis in MDD patients, which contrasts with the prominent astrogliosis that commonly occurs in patients with neurodegenerative diseases. Loss and shrinkage of astrocytes also have often been reported in mouse models of depression (Czéh et al., 2006; Rajkowska & Stockmeier, 2013; Sanacora & Banasr, 2013). That reduced astrocyte functions may contribute to depression is supported by the finding that elimination of astrocytes from the rat PFC is sufficient to induce depressive-like behaviors (Banasr & Duman, 2008). Antidepressant treatments reverse these stress-induced astroglial morphological changes to restore astroglial function, as well as reverse associated depressive-like behaviors (Rajkowska & Stockmeier, 2013; Rial et al., 2016). These findings support the current theory that astroglial dysfunction contributes to the pathophysiology of MDD and that the cellular actions of antidepressants may correct or compensate for impaired function of astrocytes (Czéh et al., 2006; Manji et al., 2003).

### 7.4    Generalization of the findings

This study confirms the therapeutic viability of an intranasally-delivered anti-inflammatory cytokine to address disease mechanisms within the brain that undergird depressive symptoms like impaired cognition. It will be important to expand on these insights and determine if other anti-inflammatory cytokines, like IL-4, have similar effects within the learned helplessness paradigm to corroborate the current findings and to more fully characterize the therapeutic potential of anti-inflammatory treatment as a general strategy to moderate depressive symptomatology. There is compelling evidence that IL-4 may have pro-cognitive, neuroprotective effects on the brain similar to IL-10 as IL-4 knockout mice display cognitive impairments (Derecki et al., 2010) and increases in both IL-10 and IL-4 concentrations in the hippocampus were shown to promote LTP (Kavanagh et al., 2004). Additionally, a determination of the relative effect sizes of other anti-inflammatory cytokines will allow anti-inflammatory treatments utilizing IL-10 to be optimized, potentially in combination with or replaced by other immunomodulatory agents well-suited to the context of stress-induced neuroinflammation. Finally, to gain a sense of the breadth of IL-10's utility to treat stress- and depression-related deficits, it will be useful to test the effect of IL-10 treatment within various types of stress-induced depressive-like behavior modalities beyond impaired learning and memory. If it can be confirmed that other anti-inflammatory cytokines can act therapeutically and/or that IL-10 demonstrates beneficial effects within depressive symptoms other than impaired learning and memory, those results would serve as a strong endorsement for the general strategy of utilizing anti-inflammatory agents as clinical tools to enhance the treatment of MDD, and would also support the inflammatory theory of depression as a conceptual framework that explains, at least in part, the etiology of some subset of depression cases.

### 7.5    Translational approach

The insights gained into the anti-inflammatory and therapeutic action of IL-10 in the context of stress and neuroinflammation represent an significant proof-of-concept for the use of strategies designed to moderate neuroinflammatory processes to improve depressive symptoms, especially impaired cognition. More extensive preclinical testing is needed in animal models to gain a more complete understanding of the cellular and biomolecular players involved in IL-10's neuroprotective and pro-cognitive action. It will be important, as well, to determine optimal dosing strategies, corresponding effect sizes, and upper bounds on IL-10 dose as a safety constraint to avoid unhealthy levels of immunosuppression. Clinical strategies specifically addressing cognitive impairments associated with MDD, which are often not remediated by current antidepressants and can persist long after mood improvement, constitute a dire unmet need in the field of psychiatry. Vortioxetine, a novel SSRI with a multimodal mechanism of action (serotonin transporter (SERT) inhibition; 5-HT1A receptor agonism; 5-HT1D, 5-HT13, 5-HT17 receptor antagonism) became the first and only drug approved by the FDA to treat cognitive deficits related to MDD in 2018 (Perini et al., 2019). However, evidence evaluating the efficacy of available treatment options in the domain of cognition in depressed patients continues to be plagued by significant limitations such as inconsistent results, a limited number of studies, and small patient samples (Rosenblat et al., 2015).

Substantial progress has been made in solidifying the theoretical conceptualization of immune system dysregulation leading to depressive symptomatology. However, the degree to which immune dysregulation plays a clinically relevant role in the neurobiology at the core of MDD remains an open question. To date, clinical trials of anti-inflammatory drugs for depression have returned encouraging but mixed results (Eyre et al., 2015; Köhler-Forsberg et al., 2019), making clear that our understanding of the neuroimmune component of depressive pathophysiology is still incomplete. Nevertheless, evidence suggests that patients with inflammation-driven depressive symptoms likely constitute a still-undefined and uncharacterized MDD subtype, with disease mechanisms highly likely to be amenable to novel anti-inflammatory strategies (Miller & Hen, 2015; Raison et al., 2013). Therefore, it is important that we capitalize on opportunities to improve our current clinical approach in order to elevate MDD treatment options which, disappointingly, are still inadequate for at least one-third of patients.

Given what is known at this point, intranasal anti-inflammatory treatment will likely only be appropriate in those patients who demonstrate abnormal elevations in neuroimmune activation. It should not be forgotten that the theoretical goal of this type of therapeutic intervention is the achievement of neuroimmune homeostatic balance, thus a potent anti-inflammatory cytokine such as IL-10 should be used judiciously so as to avoid inducing dangerous levels of immunosuppression, either with inappropriately large doses or in patients whose symptoms are not driven by overactive neuroinflammatory processes. Additionally, if the excessive neuroinflammation driving depressive symptoms is the result of a defect in the downstream pathway of IL-10, a method focused on correcting the defect, rather than introducing more IL-10, would likely show greater therapeutic promise, depending on where in IL-10's signaling cascade the rate-limiting step resides. Finally, if its effect in mice is predictive of what can be expected in humans, it is doubtful that IL-10 therapy is capable of delivering broad antidepressant action as a monotherapy. However, insofar as learning and memory is a separable but related facet within overarching depressive symptomatology, the effect of IL-10 therapy addressing learning and memory deficits has proven to be remarkable in our animal model, acting rapidly as a neuroprotective and pro-cognitive agent to fully reverse severe hippocampal-dependent impairments. For this reason, intranasal IL-10 holds promise as an adjunct to existing MDD therapeutic regimes to improve cognitive function in depressed patients experiencing impairments, and thus merits further inquiry in human subjects.

VITA

Ryan Joseph Worthen was born in Fort Lauderdale, Florida, on December 26, 1985. His parents are Brad Patrick Worthen and Paula Turner Worthen. He received his primary education at Colfax Elementary School and his secondary education at Northwest Guilford Middle School and Northwest Guilford High School. In August 2004 he entered the College of Arts & Sciences of the University of North Carolina at Chapel Hill from which he was graduated with the degree of Bachelor of Science in Biology with a Second Major in Romance Languages (Spanish) and a Minor in Chemistry in December 2008. In July 2015 he was admitted to the Graduate School of the University of Miami, where he was conferred the degree of Doctor of Philosophy in Neuroscience in December 2020.